\documentclass[12pt]{iopart}
\usepackage[T1]{fontenc}
\usepackage{graphicx}
\usepackage{palatino} 
\usepackage{mathpple} 
\usepackage{lscape} 

\usepackage{color}
\newcommand*{\parti}[2]{\frac{\partial#1}{\partial#2}}
\newcommand*{\diff}[2]{\frac{\mathrm{d}#1}{\mathrm{d}#2}}
\renewcommand*{\(}{\left(}
\renewcommand*{\)}{\right)}

\begin{document}

\title{On the correlation between fragility and stretching in
  glassforming liquids} 
\author{Kristine Niss$^1$,  C{\'e}cile Dalle-Ferrier$^1$, Gilles
  Tarjus$^2$, and Christiane Alba-Simionesco$^1$}
\address{1) Laboratoire de Chimie Physique, CNRS-UMR 8000, B{\^a}timent 349, Universit{\'e}
Paris-Sud, 91405 Orsay, France\\
2) LPTMC, CNRS-UMR 7600, Universit{\'e} Pierre \& Marie Curie,  4, Place Jussieu, 75252 Paris Cedex 05, France}
\date{\today}

\begin{abstract}
  
  We study the pressure and temperature dependences of the dielectric
  relaxation of two molecular glassforming liquids, dibutyl phtalate
  and m-toluidine. We focus on two characteristics of the slowing down
  of relaxation, the fragility associated with the temperature
  dependence and the stretching characterizing the relaxation
  function. We combine our data with data from the literature to revisit
  the proposed correlation between these two quantities. We do this in light of
  constraints that we suggest to put on the search for empirical
  correlations among properties of glassformers. In particular, argue that a
  meaningful correlation is to be looked for between stretching and
  \emph{isochoric} fragility, as both seem to be constant under isochronic
  conditions and thereby reflect the intrinsic effect of temperature.

\end{abstract}

\maketitle{}

\section{Introduction}

With the goal of better understanding the physics of glasses and of
glass formation, there has been a continuing search for empirical
correlations among various aspects of the phenomenology of
glassformers. The most distinctive feature of glass formation being
the rapid increase with decreasing temperature of the viscosity and
relaxation times, correlations have essentially been looked for
between the characteristics of the latter and other thermodynamic or
dynamic quantities. Angell coined the term ``fragility'' to describe
the non-Arrhenius temperature dependence of the viscosity or (alpha)
relaxation time and the associated change of slope on an Arrhenius
plot \cite{Angell84}. He noticed the correlation between fragility and
amplitude of the heat-capacity jump at the glass transition. Earlier,
the Adam-Gibbs approach was a way to rationalize the correlation
between the viscosity increase and the configurational or excess
entropy decrease as one lowers the temperature \cite{adam65}. Since
then, a large number of empirical correlations between ``fragility''
and other properties of the liquid or of the glass have been found:
for instance, larger fragility (\emph{i.e.}, stronger deviation from
Arrhenius behavior) has been associated with (i) a stronger deviation
of the relaxation functions from an exponential dependence on time (a
more important ``stretching'') \cite{bohmer93}, (ii) a lower relative
intensity of the boson peak \cite{sokolov93}, (iii) a larger mean
square displacement at $T_g$ \cite{ngai00}, (iv) a smaller ratio of
elastic to inelastic signal in the X-ray Brillouin-spectra
\cite{scopigno03},  (v) a larger Poisson ratio \cite{novikov04} and
(vi) a stronger temperature dependence of the elastic shear modulus,
$G_\infty$, in the viscous liquid \cite{dyre06}.

For useful as they may be to put
constraints on proposed models and theories of the glass transition,
such correlations can also be misleading by suggesting causality
relations where there are no such things. It seems therefore important
to assess the robustness of empirically established correlations. In
this respect, we would like to emphasize a number of points that are
most often overlooked:

1) Fragility involves a
variation with temperature that \emph{a priori} depends on the thermodynamic
path chosen, namely constant pressure (isobaric) versus constant
density (isochoric) conditions. On the other
hand, many quantities that have been correlated to fragility only
depend on the thermodynamic state at which they are considered: this
is not the case for the variation of the excess entropy or of the
shear modulus, nor for the jump in
heat capacity measured in differential scanning calorimetry, which are 
all path dependent, but the other properties are measured either at Tg, the 
glass-transition temperature, or in the glass, where they also relate to properties 
of the liquid as it falls out of equilibrium at Tg (there may
be a residual path dependence due to the nonequilibrium nature of the
glass, but it is quite different from that occuring in the liquid). \emph{Which 
fragility then, isobaric or isochoric, should best be used in searching for
correlations ?}

2) The quantities entering in the proposed correlations are virtually
always considered at Tg. This is the case for the commonly used
measure of fragility, the ``steepness index'', which is defined as
the slope of the temperature dependence of the alpha-relaxation time on an Arrhenius 
plot with T scaled by Tg \cite{richert98}. Tg is of course only operationally defined as the
point at which the alpha-relaxation time (or the viscosity) reaches a
given value, say 100 seconds for dielectric relaxation. The
correlated properties are thus considered at a given relaxation time
or viscosity. \emph{What is the fate of the proposed correlations when one
studies a different value of the relaxation time ?}

3) Almost invariably, comparisons involve properties measured at
atmospheric pressure, for which the largest amount of data is
available. Since, as discussed in the preceding point, the properties
are also considered at a given relaxation time, an obvious
generalization consists in studying the validity of the reported
correlations under ``isochronic'' (\emph{i.e.}, constant relaxation
time ) conditions, by varying the control parameters such that the
relaxation time stays constant. \emph{How robust are then the correlations
when one varies, say, the pressure along an isochrone ?} 
In light of the above, our contention is that any putative correlation
between fragility and another property should be tested, as far as
possible, by varying the reference relaxation time, by varying the
thermodynamic state along a given isochrone, and by changing the
thermodynamic path along which variations, such as that defining the
fragility, are measured

 A better solution would certainly be to
correlate ``intrinsic'' properties of glassformers that do not depend
on the chosen state point or relaxation time. A step toward defining such an ``intrinsic'' 
fragility has been made when it was realized
that the temperature and the density dependences of the
alpha-relaxation time and viscosity of a given liquid could be
reduced to the dependence on a single scaling variable, $X=e(\rho)/T$,
with $e(\rho)$ an effective activation energy characteristic of the
high-temperature liquid \cite{alba02,tarjus04}. Evidence is merely 
empirical and is supported by the work of several groups for
a variety of glassforming liquids and polymers \cite{alba02,tarjus04,casalini04,roland05,dreyfus04,reiser05,floudas06}. 
The direct consequence of this finding is that the
fragility of a liquid defined along an isochoric path is independent
of density: the isochoric fragility is thus an intrinsic property,
contrary to the isobaric fragility. Although one could devise ways to
characterize the isochoric fragility in a truly intrinsic manner, 
independently of the relaxation time, the common measure through the
steepness index (see above) still depends on the chosen isochrone. In
looking for meaningful correlations to this isochoric steepness index, it is
clear however that one should discard  quantities that vary with pressure (or
equivalently with temperature) under isochronic conditions. As we
further elaborate in this article, the stretching parameter characterizing
the shape of the relaxation function (or spectrum) is \emph{a priori} a valid
candidate, as there is some experimental evidence that it does not
vary with pressure along isochrones \cite{ngai05a}.

The aim of the present work is to use the knowledge about pressure
and temperature dependences of the liquid dynamics to test the robustness
of proposed correlations between fragility and other properties. This
is a continuation of the work presented in reference
\cite{niss06}, where the focus was mainly on correlations
between fragility of the liquid and properties of the associated glass.
In this paper we specifically consider the correlation between fragility
and stretching. The reported correlation between the two is indeed one of the
bases of the common belief that both fragility and stretching are
signatures of the cooperativity of the liquid dynamics.

We present new dielectric spectroscopy data on the pressure dependence
of the alpha relaxation of two molecular glassforming liquids, dibutyl
phthalate (DBP) and m-toluidine.
We express the alpha-relaxation time as a function of the scaling variable $X=e(\rho)/T$ and
evaluate the density dependence of $e(\rho)$ as well as the isochoric
fragility.  We also study the spectral shape and its pressure
dependence along isochronic lines. We spend some time discussing the
methodological aspects of the evaluation of the fragility and of the
stretching from experimental data, as well as that of the conversion
from $P,T$ to $P,\rho$ data. This provides an estimate of the error
bars that one should consider when studying correlations.
Finally, by combining our data with literature data we discuss the
robustness of the correlation between fragility and stretching along
the lines sketched above.

The paper is structured as follows. Section \ref{sec:back} introduces
some concepts and earlier developments that are central for the
discussion. In section \ref{sec:exp} we present the experimental
technique. Section \ref{sec:relax} is devoted to the pressure,
temperature and density dependence of the relaxation time. In section
\ref{sec:spec} we analyze the spectral shape and its pressure and
temperature dependence. Finally, in section \ref{sec:disc} we combine
the current results with literature data to assess the relation
between fragility and stretching, stressing the need to disentangle temperature and density effects. 
Two appendices discuss some methodological points. 

\section{Background\label{sec:back}}

\subsection{Isochoric and isobaric fragilities}\label{sec:iso}

The fragility is a measure of how much the temperature
dependence of the alpha-relaxation time (or alternatively the shear
viscosity) deviates from an Arrhenius form as the liquid approaches the glass
transition. The most commonly used criterion is the so called
steepness index,
%original def. of fragility
\begin{eqnarray}
  \label{eq:angel}
  m_P=\left.\parti{\log_{10}(\tau_\alpha)}{\,T_g/T}\right|_P (T=T_g),
\end{eqnarray}
where the derivative is evaluated at $T_g$ and $\tau_\alpha$ is
expressed in seconds.  Conventionally, the liquid is
referred to as strong if $m$ is small, that is $17-30$, and fragile if
$m$ is large, meaning roughly above 60. In the original classification
of fragility it was implicitly assumed that the relaxation time (or
viscosity) was monitored at constant (atmospheric) pressure, as this is
how the vast majority of experiments are performed. The conventional
fragility is therefore the (atmospheric pressure) isobaric fragility, and, as
indicated in Eq. \ref{eq:angel}, the associated steepness index is
evaluated at constant pressure. However, the relaxation time can also
be measured as a function of temperature along other isobars, and this
will generally lead to a change in $m_P$. Moreover, it is
possible to define an isochoric fragility and the associated index,
$m_\rho$, obtained by taking the
derivative at constant volume rather than at constant pressure. The
two fragilities are straightforwardly related via the chain rule of
differentiation,
\begin{eqnarray*}
    m_P = m_\rho +\left.\parti{\log_{10}(\tau_\alpha)}{\rho}\right|_T \left.\parti{\rho}{\,T_g/T}\right|_{P}(T=T_g),
\end{eqnarray*}
when both are evaluated at the same point $(T_g(P),\rho(P,T_g(P)))$.
The isochoric fragility, $m_\rho$, describes the intrinsic effect of
temperature, while the second term on the right hand side incorporates
the effect due to the change of density driven by temperature under isobaric
conditions. It can be shown that the above relation can be rewritten as
%
%decomposition of fragility
\begin{eqnarray}
  m_P = m_\rho (1-\alpha_P/\alpha_\tau)
\label{eq:mpmrho}
\end{eqnarray}
where the unconventional $\alpha_{\tau}$ is the isochronic expansivity \cite{ferrer98},
\emph{i.e.}, the expansivity along a line of constant alpha-relaxation time
$\tau_{\alpha}$ (the $T_g$ line being a specific isochrone). 
The above result is purely formal and contains no assumptions. The
implication of the result is that $m_P$ is larger than $m_\rho$ if
$\alpha_P>0$ and $\alpha_\tau<0$. It is well known that $\alpha_P>0$
 in general. The fact that $\alpha_\tau$ is negative
arises from the empirical result that the liquid volume always decreases
when heating while following an isochrone. 

Within the last decade a substantial amount of relaxation-time and
viscosity data has been collected at different temperatures and
pressures/densities. On
the basis of the existing data, it is reasonably well established that
the temperature and density dependences of the alpha-relaxation time can be
expressed in a scaling form as \cite{alba02,tarjus04,casalini04,roland05,dreyfus04,reiser05,floudas06}.
%scaling
\begin{eqnarray}
  \tau_\alpha(\rho,T)=F\left(\frac{e(\rho)}{T}\right).
\label{eq:scaling}
\end{eqnarray}

It is seen directly from Eq. \ref{eq:scaling} that
$X(\rho,T)=e(\rho)/T$, when evaluated at $T_g$, has the same
value at all densities ($X_g=e(\rho)/T_g(\rho)$) if $T_g(\rho)$ is
defined as the temperature where the relaxation time has a given
value (e.g., $\tau_\alpha=100$ s). Exploiting this fact, it is easy to
show \cite{tarjus04,alba06} that the scaling law implies
that the isochoric fragility is independent of density. For instance,
the isochoric steepness index, when
evaluated at a $T_g$ corresponding to a fixed relaxation time, is given by

\begin{equation}
 m_\rho=\left.\diff{\log_{10}(\tau_\alpha)}{\,T_g/T}\right|_{\rho}(T=T_g)=F^\prime (X_g) \diff{X}{T_g/T}(T=T_g) = X_g F^\prime (X_g).
\end{equation}

The fact that the relaxation time $\tau_\alpha$ is constant when $X$ is
constant means that the isochronic expansion coefficient 
$\alpha_\tau$ is equal to the  expansion coefficient at constant
$X$. Using this and the general result $\left(\parti{\rho}{T}
\right)_X\left(\parti{X}{\rho} \right)_T\left(\parti{T}{X} \right)_\rho=-1$,
it follows that 

\begin{equation}
  \frac{1}{\alpha_\tau}=-T_g \diff{ \log e(\rho)}{\log \rho},
\end{equation}

which inserted in Eq. \ref{eq:mpmrho} leads to 

\begin{eqnarray}
  m_P = m_\rho \(1+\alpha_P T_g \diff{ \log e(\rho)}{\log \rho}\),
\label{eq:mpmrho2}
\end{eqnarray}
where $m_P$, $m_\rho$ and $\alpha_P$ are evaluated at $T_g$.

When liquids have different isobaric fragilities, it can be
thought of as due to two reasons: a difference in the intrinsic
isochoric fragility, $m_\rho$, or a difference in the relative effect of density,
characterized by $\alpha_P T_g$ and the parameter $x=\diff{ \log
  e(\rho)}{\log \rho}$. We analyze the data in this frame.

\subsection{Relaxation-time dependent fragility}\label{sec:time}

The following considerations hold for isochoric and isobaric
conditions alike. The $\rho$ or $P$ subscript are therefore omitted in
this section. 

The fragility is usually characterized by a criterion evaluated at
$T_g$, \emph{i. e.}, the temperature at which the relaxation time reaches
$\tau_\alpha$=100 s-1000 s. The same criterion, e.g. the steepness index, can
however equally well be evaluated at a temperature corresponding to
another relaxation time, and this is also found more often
in literature, mainly to avoid the extrapolation to long times. So
defined, the  ``fragility'' for a given system can be considered as a quantity
which is dependent of the relaxation time at which it is evaluated:
\begin{eqnarray}
  m(\tau)=\diff{log_{10}(\tau_\alpha)}{T_\tau/T} (T=T_\tau)\,
\label{eq:mtau}
\end{eqnarray}
where $\tau_\alpha(T_\tau)=\tau$ defines the temperature $T_\tau$. ($T_g$ is
a special case with $\tau\approx $100 s-1000s .)

An (extreme) strong system is a system for which the relaxation time 
has an Arrhenius behavior,
\begin{eqnarray}
  \tau_\alpha(T)=\tau_\infty\exp\(\frac{E_\infty}{T}\), 
\end{eqnarray}
where $E_\infty$ is a temperature and density independent activation energy
(measured in units of temperature). Inserting this in the expression
for the relaxation-time dependent steepness index (Eq. \ref{eq:mtau}) gives 
\begin{eqnarray}
  m_{strong}(\tau)=log_{10}\(\tau/\tau_\infty\)
\end{eqnarray}
which gives the value $m_{strong}(\tau=100s)=15$ (assuming
$log_{10}(\tau_\infty/\mathrm{sec})=-13$) and decreases to $m_{strong}(\tau=\tau_\infty)=0$
as the relaxation time is decreased. This means that even for a strong
system the steepness index is relaxation-time dependent. In order to get
a proper measure of departure from Arrhenius behavior it could
therefore be more adequate to use the steepness index normalized by that of a strong system:

\begin{eqnarray}
  m_n(\tau)=\frac{ m(\tau)}{ m_{strong}(\tau)}=\frac{\diff{log_{10}(\tau_\alpha)}{T_\tau/T}}{log_{10}\(\tau/\tau_\infty\)}.
\label{eq:mtaun}
\end{eqnarray}
$m_n(\tau)$ will take the value $1$ at all relaxation times in a
system where the relaxation time has an Arrhenius behavior. Such a normalized
measure of fragility has been suggested before
\cite{schug98,granato02,dyre04}. For instance, Olsen and
coworkers \cite{dyre04} have introduced the index

\begin{eqnarray}
  I=-\diff{log E(T)}{log T}
\end{eqnarray}

where $E(T)$ is a temperature dependent activation energy defined by
$E(T)=T\ln{(\tau_\alpha/\tau_infty)}$. The relation between the steepness index and
the Olsen index is \cite{dyre04} 

\begin{equation}
  \label{eq:mI2}
  I(\tau)=\frac{m(\tau)}{\log_{10}\(\frac{\tau}{\tau_\infty}\)}-1=m_n(\tau)-1
\end{equation}

$I(\tau)$ takes the value 0 for strong systems at all relaxation
times. Typical glass forming liquids display an approximate Arrhenius behavior at
high temperatures and short relaxation times; in this limit
$I(\tau)=0$ and it increases as the temperature dependence starts
departing from the Arrhenius behavior. Typical values of I at
$T_g(\tau=100s)$ are ranging from I=3 to I=8, corresponding to steepness indices of m=47 to m=127. 

Finally, we note in passing that relaxation-time independent measures
of fragility can be formulated through fitting formulae:
this is the case for instance of the fragility parameter $D$ in the
Vogel-Tammann-Fulcher (VTF) formula or of the frustration parameter $B$ in
the frustration-limited domain
theory \cite{kivelson98}.
 
\section{Experimentals}\label{sec:exp}

The dielectric cell is composed of two gold-coated electrodes
separated by small Teflon spacers. The distance between the spacers is
$0.3$ mm and the area is 5.44 cm$^{2}$ giving an empty capacitance of
16 pF. The electrodes are totally immersed in the liquid sample, which
is sealed from the outside by a Teflon cell. The electric contacts are
pinched through the Teflon. The compression is performed using liquid
pentane, which surrounds the Teflon cell from all sides. The Teflon
cell has one end with a thickness of $0.5$ mm in order to insure that
the pressure is well transmitted from the pentane to the 
liquid sample. The pressure is measured by using a strain gauge. The cooling is performed
by a flow of thermostated cooling liquid running inside the
autoclave. The temperature and the temperature stability are monitored
by two PT100 sensors placed 2cm and 0.3 cm from the sample. The
temperature is held stable within $\pm 0.1$ degree for a given
isotherm. The temperature during the time it takes to record a spectrum
is stable within $\pm 0.01$ degree. 

The setup insures a hydrostatic pressure because the sample is
compressed from all sides. It is moreover possible to take spectra
both under compression and decompression. By doing so and returning to
the same $P-T$ condition after several different thermodynamic
paths, we have verified that there was no hysteresis in the pressure
dependence of the dynamics. This serves to confirm that the liquid is
 kept at thermodynamic equilibrium at all stages.

The capacitance was measured using a HP 4284A LCR-meter which covers
the frequency range from 100 Hz to 1 MHz. The low-frequency range from
100 Hz to 1 Hz is covered using a SR830 lockin.  

The samples, dibutyl phthalate (DBP) and m-toluidine, were acquired from
Sigma-Aldrich. The m-toluidine was twice distilled before usage. The
DBP was used as acquired.

Liquid m-toluidine was measured on one isotherm at 216.4 K.  DBP was measured
along 4 different isotherms, 205.5 K, 219.3 K, 236.3 K and 253.9 K, at
pressures up to 4 kbar. DBP was moreover measured at different
temperatures along two isobars: atmospheric pressure and 230 MPa. The
pressure was continuously adjusted in order to compensate for the
decrease of pressure which follows from the contraction of the sample
due to decreasing temperature. It is of course always possible to
reconstruct isobars based on experiments performed under isotherm
conditions. However, such a procedure mostly involves interpolation of
the data, which is avoided by performing a strictly isobaric
measurement.  For DBP we have obtained relaxation-time data at times
shorter than $10^{-6.5}$ s by using the high-frequency part of the
spectrum and assuming time-temperature and time-pressure superposition
(TTPS). Although TTPS is not followed to a high precision (see 
section \ref{sec:shapeDBP}), the discrepancies lead to no significant error on the
determination of the relaxation time. This is verified
by comparison to atmospheric-pressure data from the literature (see figure
\ref{fig:dbpIsob}).

\section{Alpha-relaxation time and fragility}\label{sec:relax}

\subsection{Dibutyl phtalate}\label{sec:relaxDBP}

\begin{figure}[htbp]
  \centering
  \includegraphics[scale=0.4]{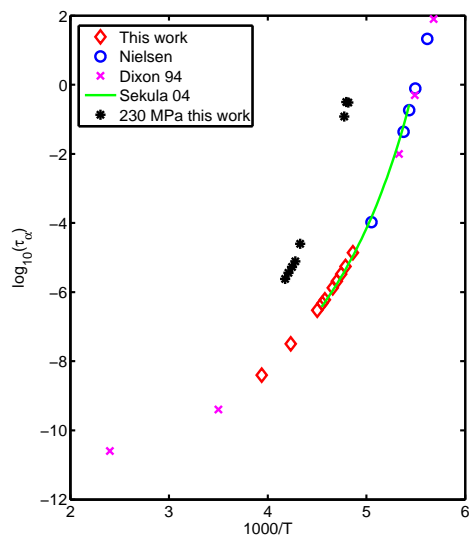}
  \caption{Temperature dependence of the alpha-relaxation time
    (from dielectric measurements, $\tau_\alpha=1/\omega_{peak}$) of liquid DBP at atmospheric pressure and at 230
    MPa (Arrhenius plot). Data at atmospheric pressure from other groups are also included: unpublished data
    from Nielsen \cite{nielsen06}, the VTF fit of \cite{sekula04} shown in the range where it can be
    considered as an interpolation of the original data, and data taken from figure 2(a) in reference
    \cite{dixon90}.}\label{fig:dbpIsob}
  \end{figure}

The DBP data at atmospheric pressure is shown in figure \ref{fig:dbpIsob} along
with literature results.  
$T_g(P_atm)=177$ K, when defined as the
temperature at which $\tau_\alpha=100$ s. We also present the data taken at
$P=230$ MPa in this figure. It is clearly seen that $T_g$ increases
with pressure. An
extrapolation of the data to $\tau_\alpha=100$ s gives $T_g=200$ K for
$P=230$ MPa,
corresponding to $d T_g/ d P\approx0.1$ KMPa$^{-1}$. This corresponds well to
the pressure dependence of $T_g$ (at $\tau_\alpha=1$ s) reported by
Sekula \emph{et al.} \cite{sekula04}, based on measurements taken at
pressures higher than $600$ MPa. The dependence is however stronger than that reported by
Fujimori \emph{et al.} \cite{fujimori97} based on isothermal
calorimetry, for which $d T_g/ d P=0.06$ KMPa$^{-1}$. This indicates that the calorimetric and the dielectric
relaxations may have somewhat different dependences on pressure. 

\begin{figure}[htbp]
  \centering
  \includegraphics[scale=0.4]{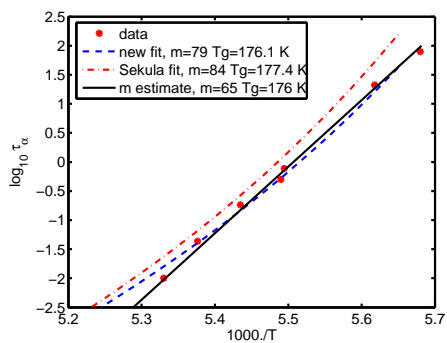}
  \caption{Atmospheric-pressure data of figure \ref{fig:dbpIsob} 
    with relaxation times longer than a millisecond (symbols). Also shown are the VTF fit from
    reference \cite{sekula04} extrapolated to low temperatures
    (dashed-dotted line), a new VTF fit made by using data in the
    $10^{-6}$ s $-10^{2}$ region (dashed line), and estimated slope of
    the data in the long-time
    region (full line). The $Tg$'s estimated from these three methods are very
    similar, whereas the fragility varies significantly from $m=65$ to
    $m=85$.  }\label{fig:fragi}
  \end{figure}

In figure \ref{fig:fragi} we illustrate the determination of $T_g$ and
of the steepness index $m_P$ for the atmospheric-pressure data, using the
part of the data of figure \ref{fig:dbpIsob} whith a
 relaxation time longer than a millisecond. Along with the data
we show the VTF fit from Sekula \emph{et al.} \cite{sekula04}
extrapolated to low temperatures, which gives
$T_g=177.4$ K and $m_P=84$. We have also performed a new VTF fit
restricted to the data in the $10^{-6}$ s$-10^{2}$ s region. The result of
this fit yields $T_g=176.1 $ K and $m_P=79$. Finally, we have made a simple
linear estimate of $log_{10}\tau_\alpha$ as a function of $1/T$ in the
temperature range shown in the figure. This linear slope fits the data
close to $T_g$ better than any of the VTF fits. The corresponding
glass transition temperature and steepness index are $T_g=176$ K and $m_P=65$.
This illustrates that the determination of $T_g$ is rather robust
while this is less so for the steepness index. This latter
depends on how it is obtained, and the use of extrapolated
VTF fits can lead to an overestimation. (Of course, a VTF fit made over
a very narrow range, e.g.  $10^{-2}s-10^{2}$ s, will agree with the
linear fit, because it becomes essentially linear over the
restricted range.) The fragility of DBP has earlier been reported to be
$m_P=69$ \cite{bohmer93}, based on the data of Dixon
\emph{et al.} \cite{dixon90}. We take $m_P=67$ as a representative value.

\begin{figure}[htbp]
  \centering
  \includegraphics[scale=0.4]{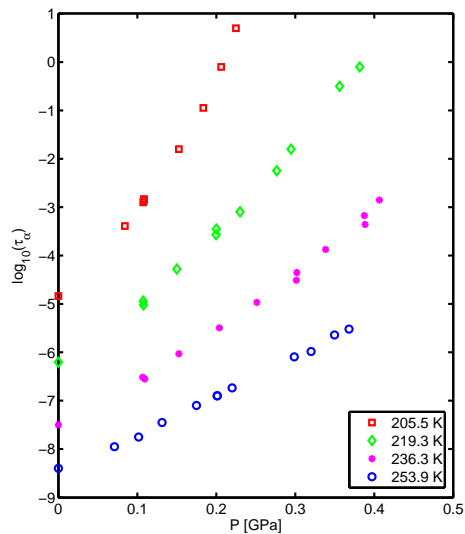}
  \caption{Alpha-relaxation time of DBP (from dielectric measurements,
    $\tau_\alpha=1/\omega_{peak}$) as a function of pressure along 4 different
    isotherms (log-linear plot).}\label{fig:dbpP}
  \end{figure}

The relaxation-time data along four different isotherms are displayed
as a function of pressure in figure \ref{fig:dbpP}.

In order to separate the relative effects of density and temperature
it is convenient to express the relaxation
time as a function of density and temperature rather than pressure and
temperature. To do this, we need the pressure and temperature dependences of
the density. However, for liquid DBP such data is only available at high
temperature \cite{bridgman32}.

In order to extrapolate the equation of state to low
temperature we have applied the following scheme. When calculated from the data in Ref. \cite{bridgman32}, the
expansion coefficient $\alpha_P$ shows a weak decrease with
decreasing temperature. We therefore assume that the temperature
dependence of $\alpha_P$ is linear over the whole temperature range
and integrate with respect to temperature to obtain the density along the atmospheric-pressure isobar. 
In the whole temperature range of Ref. \cite{bridgman32}, the
pressure dependence of the density is well described by fits to the
Tait equation with temperature-dependent adjustable parameters ``c'' and ``b''
\cite{cook93} (which are directly related to the compressibility
and its first-order pressure derivative). We have linearly
extrapolated the temperature dependence of these parameters and used the 
Tait equation to calculate the
pressure dependence along each isotherm.  Extrapolating the
derivatives rather than the density itself is expected to lead to smaller
errors on the latter. In addition, we have checked that
this procedure gives physically reasonable pressure and temperature
dependences of the expansivity and of the compressibility
\cite{terminassian88}.

\begin{figure}[htbp]
  \centering
\includegraphics[scale=0.4]{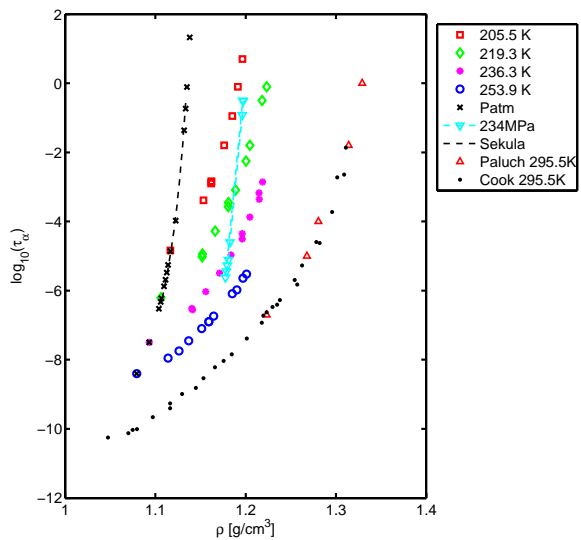}
  \caption{Logarithm of the alpha-relaxation time of DBP versus density (see the text regarding the
    calculation of density). Included are data from this work along with
    dielectric data from figure 3 in reference \cite{paluch03}, and viscosity data from reference
    \cite{cook93}. The viscosity data are shifted arbitrarily on
    the logarithmic scale in order to make the absolute values
    correspond to the dielectric data of reference
    \cite{paluch03}, which are taken at the same
    temperature.}\label{fig:dbpRho}
  \end{figure}

Figure \ref{fig:dbpRho} shows the density dependence of the alpha-relaxation
time along the four different isotherms, the atmospheric-pressure
isobar and the 230 MPa isobar. We have also included the room-temperature
dielectric data of Paluch \emph{et al.} \cite{paluch03}. For DBP the viscosity
data and the dielectric relaxation time do not decouple under pressure
\cite{sekula04}, and we have therefore also included the room-temperature
viscosity data of Cook \emph{et al.} \cite{paluch03}.

\begin{figure}[htbp]
  \centering
  \includegraphics[scale=0.4]{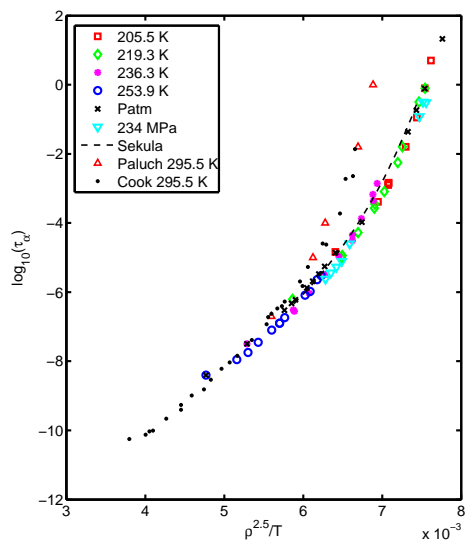}
  \caption{The alpha-relaxation times shown in figure \ref{fig:dbpRho}
  plotted as a function of $\rho^{2.5}/T$.}\label{fig:dbpScaling}
  \end{figure}

\begin{figure}[htbp]
  \centering
 a) \includegraphics[scale=0.4]{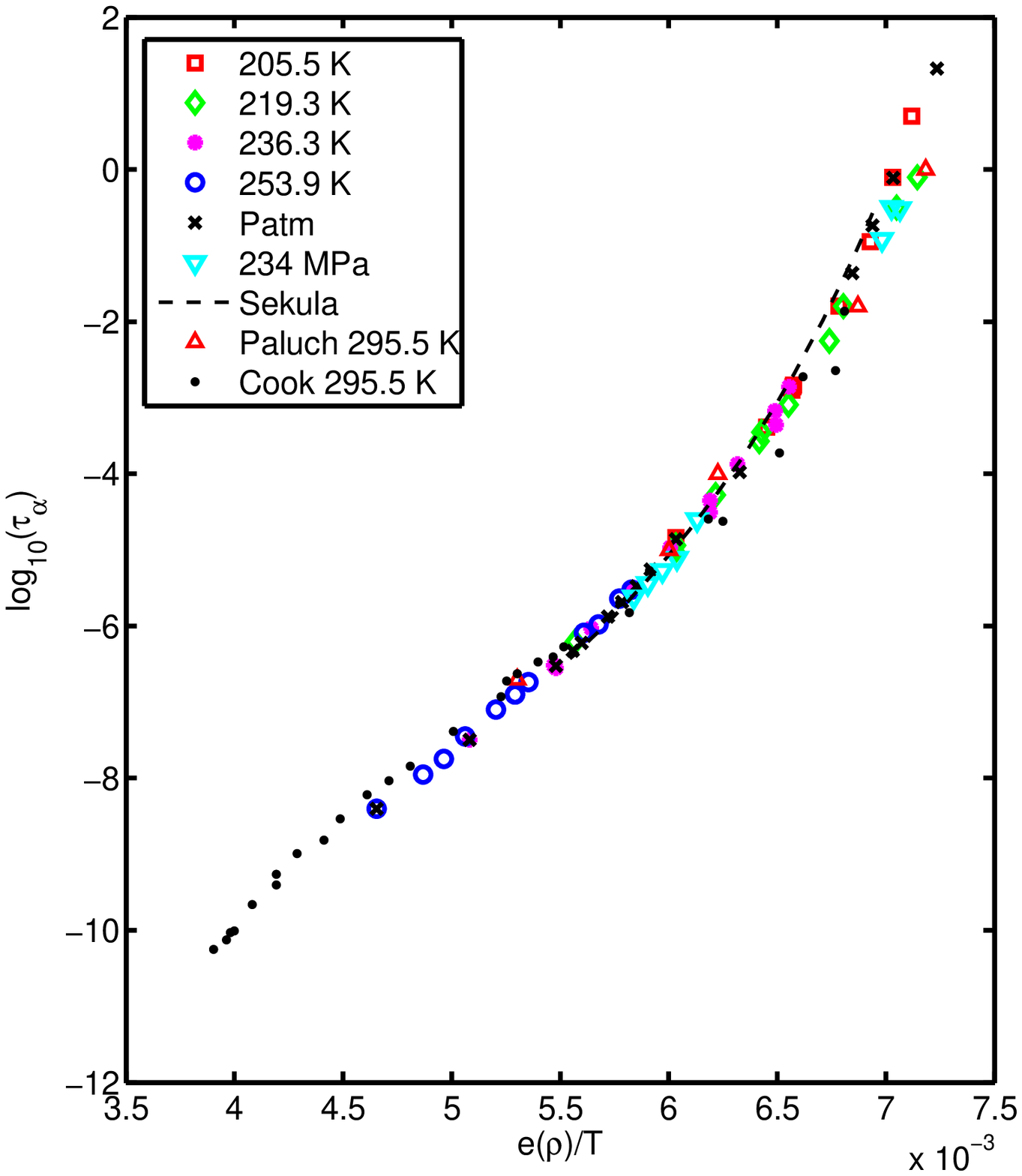}
 b) \includegraphics[scale=0.4]{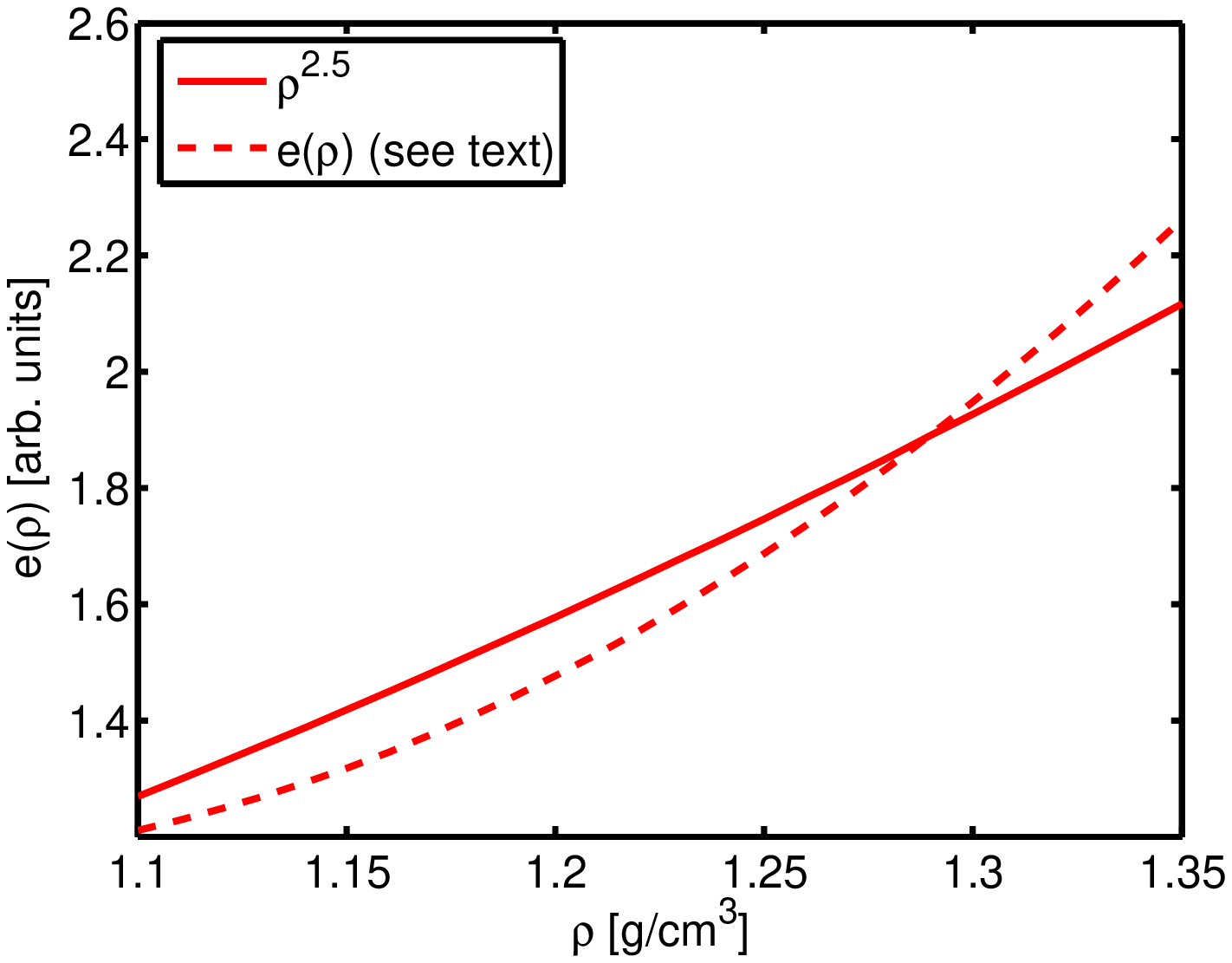}

  \caption{(a) The alpha-relaxation times shown in figure \ref{fig:dbpRho}
    plotted as a function of $X=e(\rho)/T$, with increasing $d log
    e(\rho)/d log \rho$ as $\rho$ increases. (b) Density-dependent
    activation energy $e(\rho)$ (dashed line) used in the scaling
    variable $X=e(\rho)/T$ for collapsing data in (a) (the associated
    $x(\rho)=d log e(\rho)/d log \rho$ increases from 1.5 to 3.5 in the
     density range under study). We also display
    the power law giving the best scaling, $\rho^{2.5}$, at low
    density (full line).
    }\label{fig:dbpScaling2}
  \end{figure}

In figure \ref{fig:dbpScaling} we show the data of figure
\ref{fig:dbpRho} plotted as a function of the scaling variable
$\rho^x/T$, choosing for $x$ the value that gives the best collapse for
the data of this work. This corresponds to testing the scaling in
equation \ref{eq:scaling} by assuming that $e(\rho)$ is a power law. The
data taken at low density collapse quite well with $x=2.5$, while
this is not true for the data of Paluch \cite{paluch03} taken at
densities higher than approximately 1.2 g/cm$^3$. It is possible to
make all the data collapse by allowing $e(\rho)$ to have a stronger
density dependence at higher densities. In figure
\ref{fig:dbpScaling2} we show the data as a function of $e(\rho)/T$,
where we have constructed the density dependence of $e(\rho)$ in order
to get a good overlap of all the data (we did not look for the best
collapse, but merely evaluated the change of the isochronic
expansivity: see section \ref{sec:back}). The resulting
density dependence of $e(\rho)$ is shown in figure \ref{fig:dbpScaling2}
along with the $\rho^{2.5}$ power law. Note that the quality of the
data collapse depends only on the density dependence of $e(\rho)$ not
on its absolute value. The constructed $e(\rho)$ has an  
apparent ``power-law'' exponent $x(\rho)=\mathrm{d} log e(\rho)/\mathrm{d} log \rho$ that increases from
1.5 to 3.5 with density in the range considered. In any case, the absence of collapse
in figure \ref{fig:dbpScaling} cannot be explained by errors in
estimating the PVT data: this is discussed in more detail in Appendix \ref{sec:densAp}.

As a last note regarding the $e(\rho)/T$-scaling in figure
\ref{fig:dbpScaling2}, we want to stress that we cannot test the
scaling (Eq. \ref{eq:scaling}) in the density range above $1.25
g/cm^3$ where there is only one set of data. (This is why we did not attempt to fine 
tune  $e(\rho)$ to find the best collapse, see above.) Indeed, with a unique set
of data in a given range of density it is always possible to construct
$e(\rho)$ in this range to make the data overlap with data taken
in other density ranges.

\begin{figure}[htbp]
  \centering
  \includegraphics[scale=0.4]{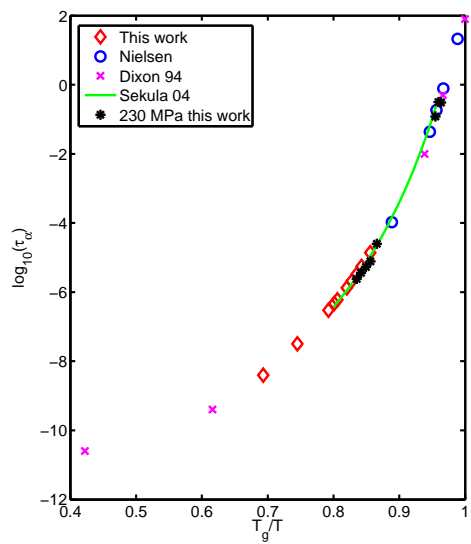}
  \caption{Arrhenius plot of the  alpha-relaxation time of DBP at
    atmospheric 
pressure and at 230 MPa, when the temperature is scaled with the
pressure dependent
    $T_g$, $T_g(Patm)=176$ K and $T_g$(230MPa)=200K. As in figure
    \ref{fig:dbpIsob}, data from other groups are also included:
    unpublished data from Nielsen \cite{nielsen06}, the VTF fit of
    \cite{sekula04} shown in the range where it can be
    considered as an interpolation of the original data, and data taken from figure 2 (a) in reference
    \cite{dixon90}.}\label{fig:dbpIsob2}
  \end{figure}

We have determined the ratio between the isochoric fragility and the isobaric
fragility at atmospheric pressure by calculating 
$\alpha_\tau$ along the isochrone of $100$ s and inserting it in Eq.
\ref{eq:mpmrho}. This leads to $m_P/m_\rho\approx 1.2$, when $m_P$. In figure \ref{fig:dbpIsob2} we show the
isobaric data taken at atmospheric pressure and at 230MPa scaled by
their respective $T_g(P)$. No
significant pressure dependence of the isobaric fragility is observed when going
from atmospheric pressure to 230 MPa, which is consistent with the result of
reference \cite{sekula04}.  The pressure independence of $m_P$ is
connected to the relatively low value of $m_P/m_\rho=1.2$ (typical
values are 1.1-2 \cite{tarjus04}); $m_\rho$ is pressure independent and the ratio
$m_P/m_\rho$ cannot be lower than one (see Eq. \ref{eq:mpmrho}),
so that $m_P$ can at most decrease by $20\%$ from its
atmospheric-pressure value. Such a change would almost be within
the errorbar of the determination of $m_P$ from the data at $230$MPa
(see the discussion earlier in this section).

\subsection{m-Toluidine}

The glass transition temperature at atmospheric pressure is $T_g=187$ K (for $\tau_\alpha=100$ s) 
and the isobaric fragility  based on dielectric spectra is reported to
be $m_P=82\pm3$ \cite{mandanici05,alba99}. (There has been some
controversy about the dielectric relaxation in m-toluidine, see reference
\cite{mandanici05} and references therein.)

\begin{figure}[htbp]
  \centering
  \includegraphics[scale=0.4]{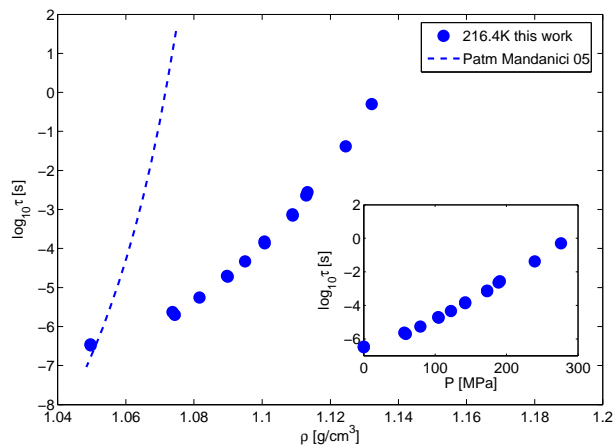}
  \caption{Logarithm of the alpha-relaxation time of m-toluidine as a function of density along
    the isotherm $T=216.4K$ (symbols). The VTF fit of the atmospheric-pressure data
    of reference \cite{mandanici05} is also shown in the range where
    the fit can be considered as an interpolation of the data (dashed line). The
    inset shows the alpha-relaxation time of
    m-toluidine as a function of pressure along the isotherm T=216.4 K.
  }\label{fig:mTrho}
  \end{figure}

In the inset of figure \ref{fig:mTrho} we show the pressure-dependent
alpha-relaxation time at $216.4$ K. Extrapolating the data to
$\tau_\alpha=100$ s leads to $P_g=340\pm10$ MPa, which is in agreement
with the slope, $d T_g/ d P=0.45$ KMaP$^{-1}$, reported for the calorimetric
glass transition in \cite{alba97}. This indicates that the
decoupling between the timescales of dipole relaxation and of
calorimetric relaxation which appears under pressure in the case of DBP
is not present in m-toluidine in this pressure range.

As for DBP, we wish to convert the temperature and pressure
dependences of the relaxation time to the temperature and density
dependences. Density data are available along four isotherms in the
$278.4$ K $-305.4$ K range for pressures up to 300 Mpa
\cite{wurflinger}.  Tait fits and thermal expansivity in this range
were extrapolated by using the scheme described above for DBP in order
to determine density both as a function of temperature down to $T_g$,
and as a function of pressure on the $216.4$ K isotherm.  In figure \ref{fig:mTrho} we show
the alpha-relaxation time as a function of density.  The data taken at
atmospheric pressure and the data taken along the 216.4 K isotherm
cover two different ranges in density. It is therefore not possible
from this data to verify the validity of the scaling in $X=e(\rho)/T$.
We therefore assume that the scaling is possible. Moreover, due to the
paucity of the data we describe $e(\rho)$ by a simple power law,
$e(\rho)=\rho^x$. We find the exponent $x$ by exploiting the fact that
the scaling variable $X=e(\rho)/T$ is uniquely fixed by the value of
the relaxation time; applying this at $T_g$, namely setting
$X_g(Patm)$ =$X_g(216$K), leads to $x=2.3$ and gives a ratio of
$m_P/m_\rho=1.2$.

\section{Spectral shape and stretching}\label{sec:spec}

The shape of the relaxation function (or spectrum), most specifically its distinctly nonexponential 
(or non-Debye) character in the viscous regime, is taken as one of the important features of glassforming materials.
Characterizing and quantifying this effect is however not fully straightforward and has led to diverging interpretations. 
First of all, the shape of the relaxation function or spectrum may change with the experimental probe considered. 
Even when restricting comparison to a single probe, here, dielectric relaxation, there is no consensus on how to 
best characterize the shape. We discuss in appendix \ref{sec:shape} various procedures that are commonly used and we 
test their validity on one representative spectrum. For reasons detailed in that appendix, we focus in the following 
on the Cole-Davidson fitting form.
  
\subsection{Dibutyl phtalate}\label{sec:shapeDBP}

\begin{figure}[htbp]
  \centering
\includegraphics[scale=0.5]{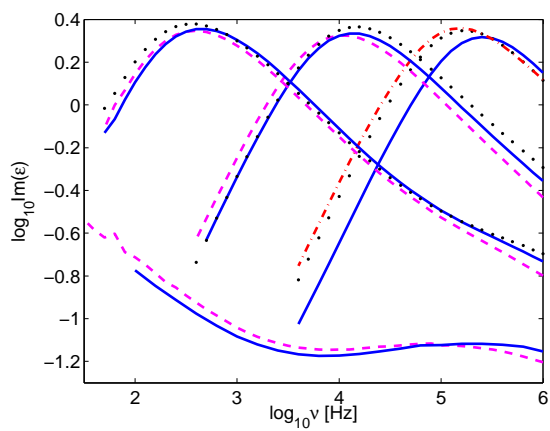}
  \caption{Log-log plot of the frequency-dependent 
dielectric loss of DBP. The curves can be sorted in 4 groups, each group having roughly the same peak frequency; from right to left: (i) Red dashed-dotted curve: T=253.9 K
    P=320 MPa; black dots: T= 236.3 K and, from right to left, P=153 MPa,
    P=251 MPa, P=389 MPa; full blue line: T=219.3 K and, from right to left,
    P=0 MPa, P=108 MPa, P=200 MPa, P=392 MPa; magenta dashed curve:
    T=206 K and, from right to left, P=0 MPa, P=85 MPa, P=206 MPa.}\label{fig:dbpimag}
  \end{figure}

\begin{figure}[htbp]
  \centering
  \includegraphics[scale=0.5]{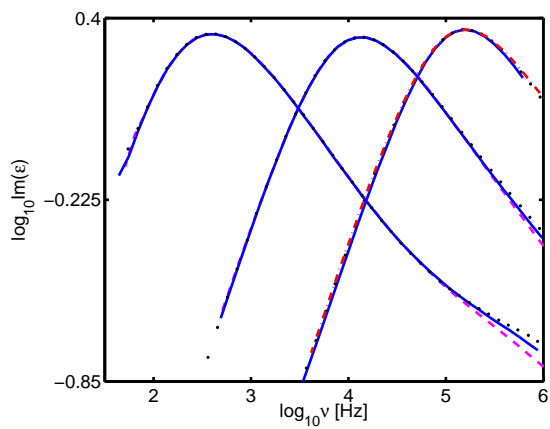}
  \caption{Same dielectric loss data  of
    DBP as in figure \ref{fig:dbpimag} with a slight shift of the peak frequencies (less than 0.2
    decade) to make the data taken under quasi
    isochronic conditions precisely coincide. The symbols are the same
    as in figure \ref{fig:dbpimag}, but the data at T=206 K and
    P=206 MPa and 219.3 K and P=392 MPa are not shown.}\label{fig:dbpimag2}
  \end{figure}

The frequency-dependent dielectric loss for a selected set of different
pressures and temperatures is shown in figure \ref{fig:dbpimag}. The
first observation is that cooling and compressing have a similar effect
as both slow down the alpha relaxation and separate the alpha
relaxation from higher-frequency beta processes. The data displayed are
chosen so that different combinations of temperature and pressure give
almost the same relaxation time. However, the correspondence is not
perfect. In figure \ref{fig:dbpimag2} we have thus slightly shifted
the data, by at most 0.2 decade, in order to make the peak positions 
overlap precisely. This allows us to compare the spectral shapes
directly. It can be seen from the figure that the shape of the alpha
peak itself is independent of pressure and temperature for a given
value of the alpha-relaxation time (\emph{i.e.}, of the frequency of the peak
maximum), while this is not true for the high-frequency part of the
spectra which is strongly influenced by the
beta-relaxation peak (or high-frequency wing). When comparing datasets
that have the same alpha-relaxation time one finds that the
high-frequency intensity is higher for the pressure-temperature
combination corresponding to high pressure and high temperature.

\begin{figure}[htbp]
  \centering
  \includegraphics[scale=0.5]{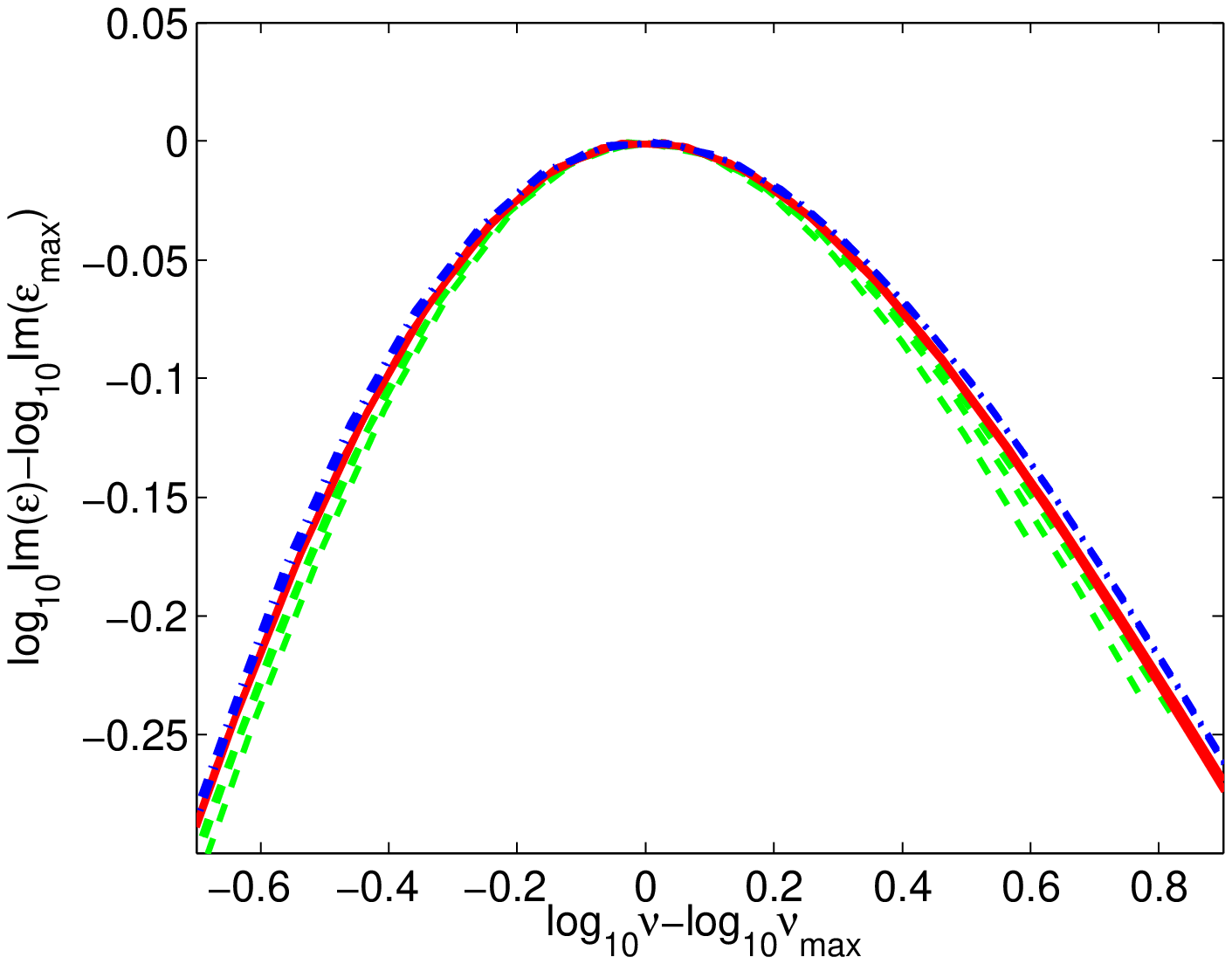}
  \caption{Same dielectric-loss data as in figures \ref{fig:dbpimag}
    and \ref{fig:dbpimag2}, with the frequency and intensity now
    scaled by the values at the maximum. We show only a $1.5$ decade
    in frequency in order to magnify the details. Notice a small
    broadening as the characteristic relaxation time increases: Blue
    dashed-dotted line are three different data sets with
    $log_{10}\nu_{max}\approx 2.6$ (P=320 MPa,T=253.9 K and
    P=153 MPa,T=236.3 K and P=0 Mpa,T=219.3 K). Red full lines are three
    data sets with $log_{10}\nu_{max}\approx 4.1$ (P=251 MPa,T=236.3 K
    and P=108 MPa,T=219.3 K and P=0 Mpa,T=205.6 K). Green dashed lines are
    three data sets with $log_{10}\nu_{max}\approx 5.2$
    (P=339 MPa,T=236.3 K and P=200 MPa,T=219.3 K and P=85 Mpa,T=205.6 K).
  }\label{fig:dbpttszoom}
  \end{figure}

In figure \ref{fig:dbpttszoom} we show all the datasets of figure
\ref{fig:dbpimag} superimposed and we zoom in on the region of the peak
maximum. The overall shape of the alpha relaxation is very similar at
all pressures and temperatures. However, looking at the data in more
detail, one finds a significantly larger degree of collapse between
spectra which have the same relaxation time, whereas a small broadening of
the alpha peak is visible as the relaxation time is increased.  At
long relaxation times there is a perfect overlap of the
alpha-relaxation peaks which have the same relaxation time.  At
shorter relaxation time, $log_{10} (\omega_{max}) \approx 5$, the
collapse is not as good: the peak gets slightly broader when pressure
and temperature are increased along the isochrone. In all cases, the
alpha peak is well described by a Cole-Davidson (CD) shape. The
$\beta_{CD}$ goes from 0.49 to 0.45 on the isochrone with shortest
relaxation time and decreases to about 0.44 close to $T_g$ at all pressures.  On the other hand, a Kolraush-William-Watts 
(KWW) fit close to $T_g$ gives $\beta_{KWW}=0.65$. A detailed discussion of the fitting procedures and of the relation between CD and KWWW descriptions is given in appendix \ref{sec:shape}.

\subsection{m-toluidine}

\begin{figure}[htbp]
  \centering
  \includegraphics[scale=0.4]{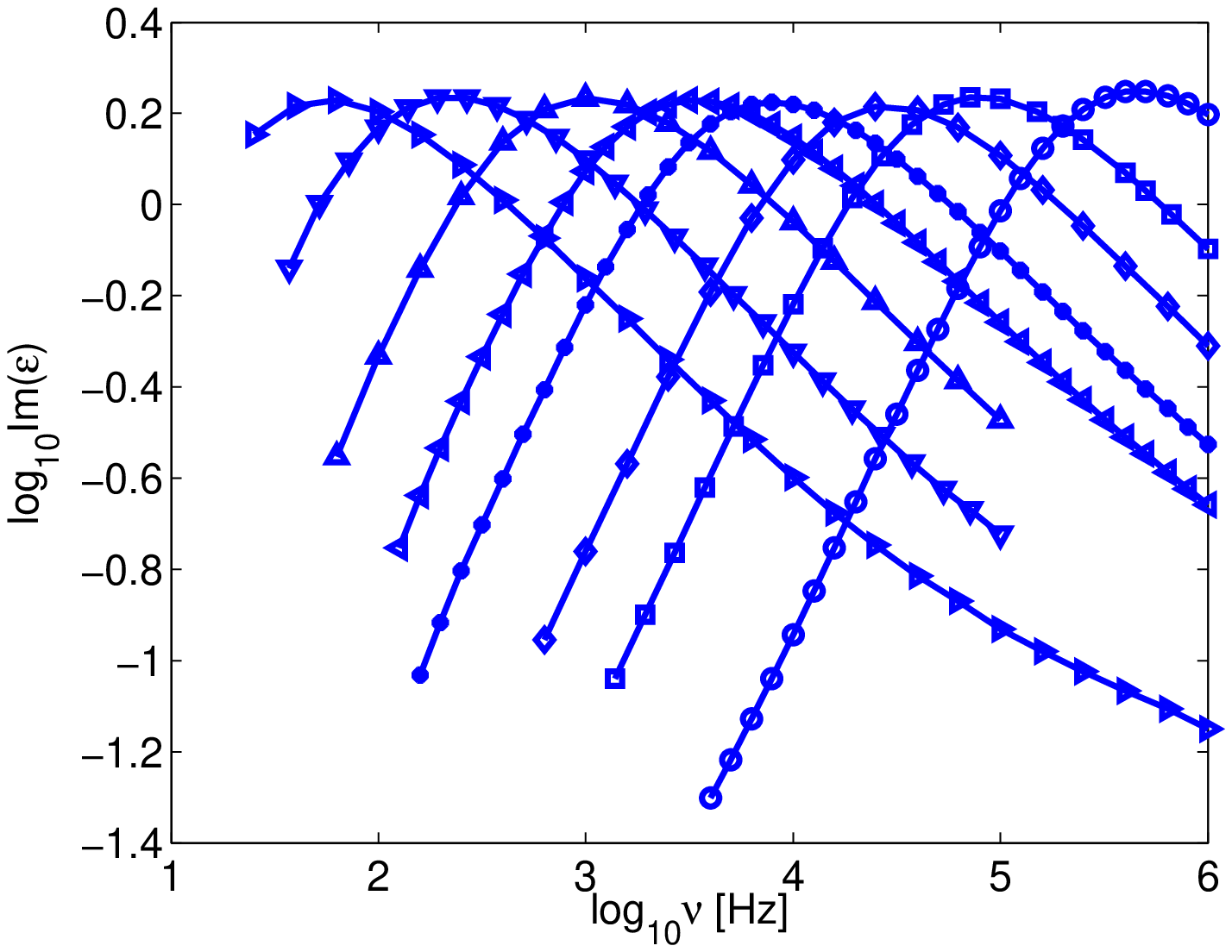}
  \caption{Log-log plot of the frequency-dependent  dielectric-loss of m-toluidine at T=216.4K and
    pressures  0 MPa, 59 MPa, 79 MPa, 105 MPa, 122 MPa, 142 MPa, 173 MPa and
    191 MPa. The peak shifts left as pressure is applied. Lines are
    guides to the eye.}\label{fig:mtolimag}
  \end{figure}

\begin{figure}[htbp]
  \centering
(a)  \includegraphics[scale=0.4]{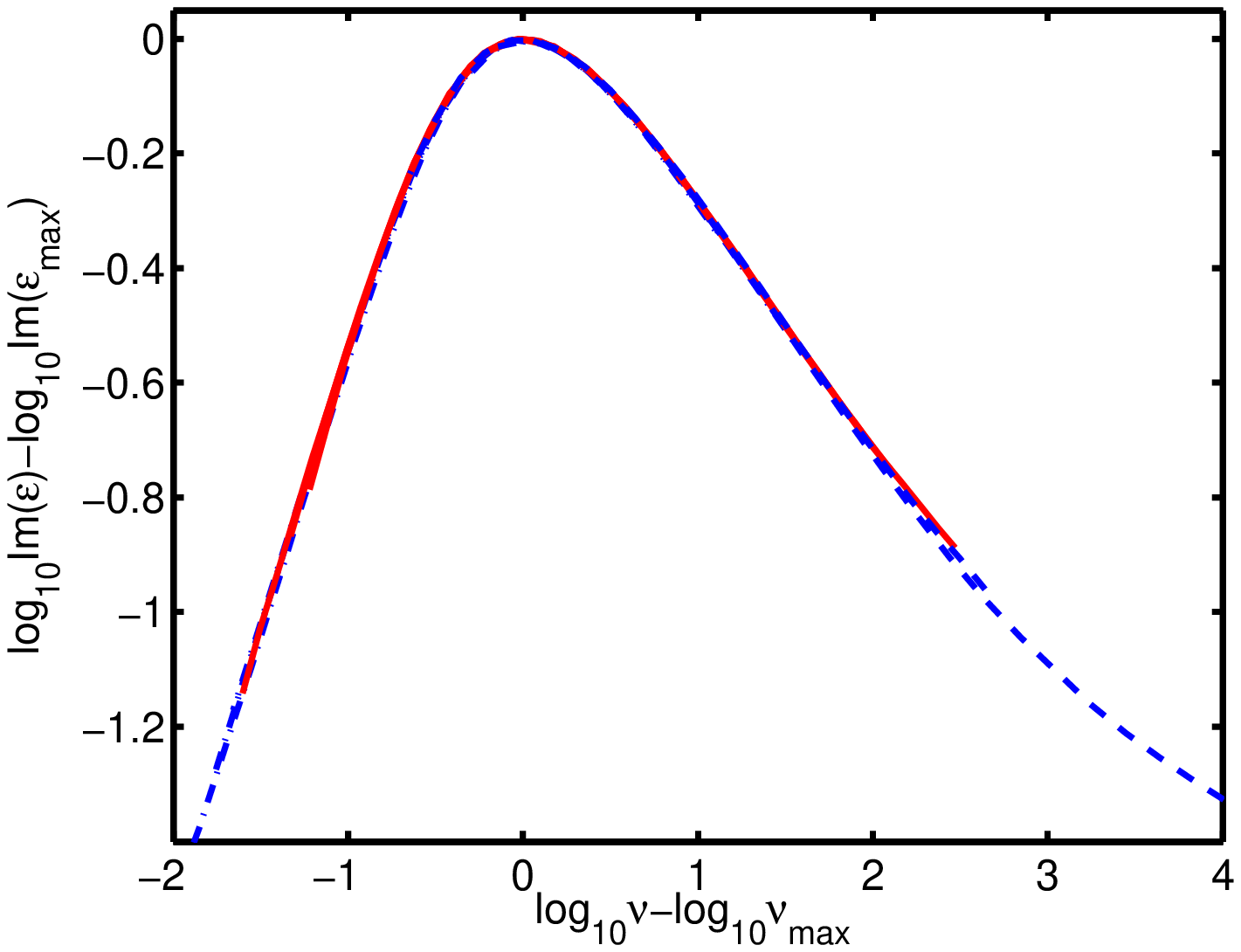}
(b)  \includegraphics[scale=0.4]{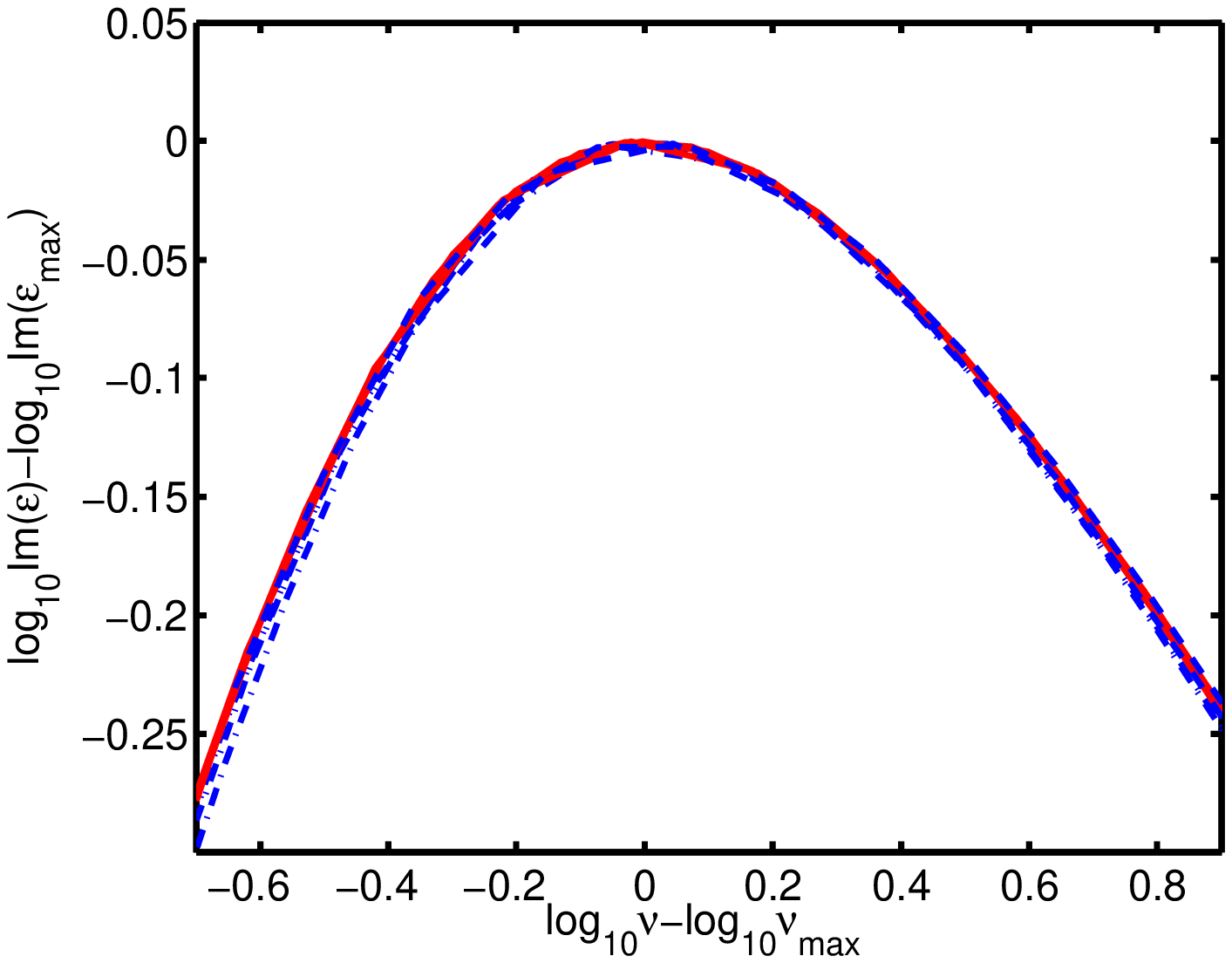}
  \caption{Same dielectric-loss data as in figure \ref{fig:mtolimag},
    now with the intensity and the frequency scaled by the values of the
    peak maximum. Figure (b) shows a zooming in of the data in (a) to
    focus on the alpha-relaxation region near the peak maximum.}\label{fig:mtoltts}
 \end{figure}

The frequency-dependent dielectric loss of m-toluidine for several
pressures along the T=216.4 K isotherm is shown in figure
\ref{fig:mtolimag}.  The data are then superimposed by scaling the
intensity and the frequency by the intensity and the frequency of the
peak maximum, respectively: this is displayed in figure
\ref{fig:mtoltts}. When zooming in (figure \ref{fig:mtoltts} (b) we
still see almost no variation of the peak shape. For the present set
of data, pressure-time-superposition is thus obeyed to a higher
degree than in DBP, and the changes are too small to give any pressure
dependence in the parameters when fitting the spectra. The
Cole-Davidson 
fit to the m-toluidine gives $\beta_{CD}=0.42$ (see also appendix 
\ref{sec:shape}). Mandanici \cite{mandanici05} and coworkers have reported
a temperature independent value of $\beta_{CD}=0.45$ for data taken at
atmospheric pressure in the temperature range 190 K-215 K, a value that
is compatible with ours. 

\section{Discussion}\label{sec:disc}

\subsection{Correlations with fragility}

As discussed in the Introduction, the temperature dependence of the alpha-relaxation
time (or of the viscosity) is usually considered as the most important phenomenon to
understand in glass science. Isobaric fragility is then often used to
characterize the viscous slowing down and its measures, such as the
steepness index, are then considered as fundamental parameters. Many
studies have been aimed at investigating which other properties of the
liquid and of the associated glass correlate to fragility. Such
correlations have been empirically established by comparing rather large sets
of systems covering a wide spectrum of fragilities.

In the literature, the finding of a correlation between fragility and some other property 
is always interpreted as indicating
that the property in question is related to the effect of \emph{temperature} 
on the structural relaxation. However, when cooling a liquid
isobarically two effects contribute to the slowing down of the
dynamics: the decrease of temperature and the associated increase of
density. Hence, the isobaric fragility is a combined measure of the
two effects. It is of course the underlying goal that the proposed correlations 
be used as guidelines and tests
in the development of theories and models for the glass transition. It
is therefore important to clarify if the correlations result from, and
consequently unveil information on, the intrinsic effect of temperature on the relaxation
time, the effect of density, or a balanced combination of the two.
 
Eq.s \ref{eq:mpmrho2} and \ref{eq:mpmrho} show how isobaric
fragility can be decomposed into two contributions, that of
temperature being given by $m_\rho$ and the relative effect of density on
relaxation time characterized by $\alpha_P T_g\frac{d log e(\rho)}{d log \rho}$. Isobaric measurements 
do not give access to $m_\rho$ nor to $\alpha_P T_g\frac{d log e(\rho)}{d log \rho}$ independently, but the relevant
information can be obtained from data taken under pressure, as we have
shown for the data presented here. From this information it becomes
possible to revisit the correlations between fragility and other
properties \cite{niss06}.  The underlying idea is that a property
supposed to correlate to the effect of temperature on the relaxation time should more specifically
correlate to the isochoric fragility, $m_\rho$, than to the isobaric one, $m_P$.

As also stressed in the Introduction, it is instructive to consider
the evolution of the empirically established correlations with
pressure. As shown in section \ref{sec:iso}, $m_\rho$ is constant,
\emph{i.e.}, is independent of density and pressure, when it is
evaluated at a pressure (or density) dependent $T_g$ corresponding to
a given relaxation time.  Nonetheless, it follows from Eq.
\ref{eq:mpmrho2}, that the isobaric fragility will in general change
due to the pressure dependence of $ \alpha_P T_g \frac{d log
  e(\rho)}{d log \rho}$. $T_g$ increases with pressure, $\alpha_P
T_g(P)$ decreases, whereas $\frac{d log e(\rho)}{d log \rho}=x$ is
often to a good approximation constant (the DBP case at high pressure
discussed in section \ref{sec:relaxDBP} is one exception). As a
result, the pressure dependence of $m_P$ is nontrivial. DBP, which we
have studied here, shows no significant pressure dependence of the
isobaric fragility, while the general behavior seen from the data
compiled by Roland \emph{et al.} \cite{roland05} is that the isobaric
fragility decreases or stays constant with pressure, with few
exceptions.  This seems to indicate that the decrease of
$\alpha_PT_g(P)$ usually dominates over the other factors.

The properties that are correlated to fragility will \emph{a priori} also
depend on pressure or density. However if a property is related to the pure effect
of temperature on the relaxation time, and therefore correlates to
$m_\rho$, then it should be independent of density when evaluated along an
isochrone (usually the glass transition line Tg), as
$m_\rho$ itself does not depend on density.

\subsection{Stretching and fragility}\label{sec:betam}
One of the properties that has been suggested to correlate to the
fragility is the nonexponential character of the relaxation function, usually expressed
in terms of the stretching parameter $\beta_{KWW}$. 

\begin{figure}[htbp]
  \centering
  \includegraphics[scale=0.4]{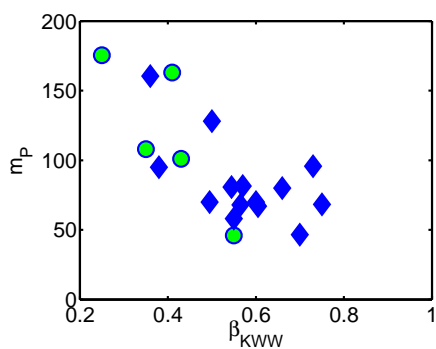}
  \caption{Isobaric fragility as a function of stretching
    parameter. Diamonds: molecular liquids, circles:
    polymers. See table \ref{table} for numerical values and references. 
  }\label{fig:mP}
  \end{figure}

\begin{figure}[htbp]
  \centering
 \includegraphics[scale=0.4]{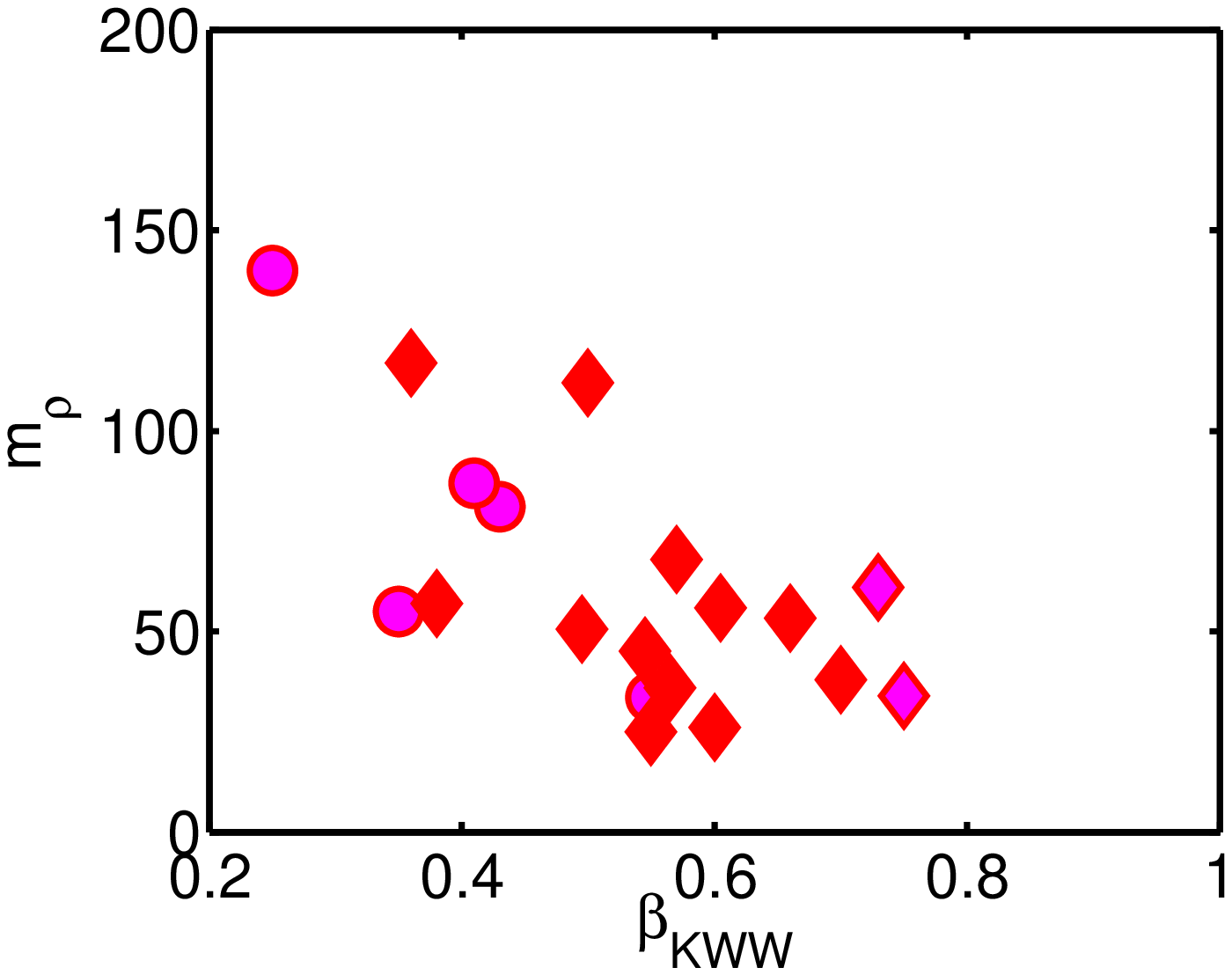}
  \caption{Isochoric fragility $m_\rho$ as a function of stretching
    parameter.
 Diamonds: molecular liquids, circles:
    polymers. See table \ref{table} for numerical values and references. 
}\label{fig:mrho}
  \end{figure}

\begin{figure}[htbp]
  \centering
  \includegraphics[scale=0.4]{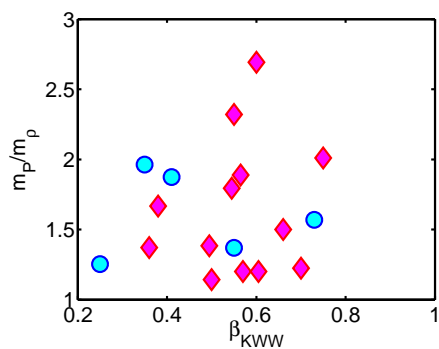}
  \caption{
    Ratio between isochoric and isobaric fragility as a function of
    stretching parameter. 
 Diamonds: molecular liquids, circles:
    polymers. See table \ref{table} for numerical values and references. 
}\label{fig:ratio}
  \end{figure}

The data we have reported here confirm the earlier finding
\cite{ngai05a} that the spectral shape of the alpha relaxation does
not vary when pressure is increased while keeping the relaxation time
constant. This leads us to suggest that, if a correlation between
fragility and stretching does exist, this latter should better
correlate to the isochoric fragility which is also independent of
pressure than to the isobaric fragility. To test this hypothesis we
have collected data from literature reporting isobaric fragility and
stretching of the relaxation at $T_g$. We consider here a description
of the shape of the relaxation function in terms of the KWW stretching
parameter $\beta_{KWW}$. This choice is made because it is convenient
to use a characterization with only one parameter for the shape (see
appendix \ref{sec:shape} for a discussion and the connection with the
Cole-Davidson description used above) and because $\beta_{KWW}$ is the
most reported of the liquids where $m_\rho$ is also available.  The
compilation of this data is shown in table I and in figures
\ref{fig:mP} and \ref{fig:mrho} where both the isobaric fragility at
atmospheric pressure (Fig. \ref{fig:mP}) and isochoric fragility (Fig.
\ref{fig:mrho}) are plotted against the stretching parameter. There is a
great deal of scatter in both figures. There is however an observable
trend, the fragilities appearing to decrease as the stretching
increases. The relative effect of density (over that of temperature)
 on the slowing down of the relaxation is characterized by the term $
\alpha_P T_g \frac{\mathrm{d} log e(\rho)}{\mathrm{d} log \rho}=m_P/m_\rho-1$.  In
figure \ref{fig:ratio} we show the ratio $m_P/m_\rho$ as a function of
$\beta_{KWW}$. Clearly, no correlation is found between this ratio and
the stretching.

\begin{landscape}
\begin{table}\label{table}
\begin{tabular}{|c|c|c|c|c|c|c|}
  \hline
  Compound & $m_{P}$ & Refs. & $m_{\rho}$ & Refs. & $\beta_{KWW}$ & Refs. \\
  \hline
  \emph{o}-terphenyl & 82, 81, 76, 84 & \cite{alba04,dixon88,huang01,paluch01} & 45 & \cite{alba04} & 0.57, 0.52 & \cite{dixon88,tolle01} \\
{\small Dibutyl phtalate} & 67 & this work & 56 &   this work  &
  0.56,0.65 & \cite{dixon90} this work \\
  PC & 104, 93, 90 & \cite{qin06,richert03,paluch01}& 57, 65$^*$  & \cite{casalini05b,reiser05} & 0.73 & \cite{paluch01} \\
  BMPC & 70 & \cite{casalini05c} & 26 & \cite{casalini05c} & 0.6 & \cite{hensel02b}\\
  BMMPC & 58 & \cite{casalini05b} & 25 & \cite{casalini05b} & 0.55 & \cite{casalini03} \\
  DEP A& 95 & \cite{roland04} & 57 & \cite{roland04} & 0.38 & \cite{paluch03b} \\
  KDE & 64,73,68 & \cite{casalini05b,paluch01,roland03} & 34 & \cite{casalini05b}& 0.75 & \cite{paluch01} \\
  DHIQ & 163, 158 & \cite{casalini06,richert03} & 117 & \cite{casalini06} &  0.36 & \cite{richert03} \\
  Cumene & 80$^*$ & \cite{barlow66} & 53$^*$& \cite{barlow66,bridgman49}  & 0.66 & \cite{nissU} \\
  Salol & 68, 73, 63 & \cite{roland05,paluch01,laughlin72} & 36 & \cite{roland05} & 0.6, 0.53  & \cite{sidebottom89,bohmer93} \\
  Glycerol & 40, 53 & \cite{alba04,birge86} & 38 &  \cite{alba04} & 0.65, 0.7 ,0.75 & \cite{birge86,ngai90,dixon90} \\
  Sorbitol & 128 & \cite{casalini04} & 112 & \cite{casalini04} & 0.5 & \cite{ngai91} \\
  \textit{m}-fluoroaniline & 70 & \cite{alba99} & 51$^*$ & \cite{reiser05} & 0.35, 0.64 & \cite{cutroni94,hensel05} \\
  \textit{m}-toluidine & 84,79 & \cite{mandanici05,alba99}& 68 & this work & 0.57 & this work \\
  Polyisobutylene & 46 & \cite{plazek91} & 34$^*$ & \cite{audePHD} & 0.55 & \cite{plazek91} \\
  Polyvinylchloride & 160, 191 & \cite{huang02,plazek91} & 140 & \cite{huang02} & 0.25 & \cite{plazek91} \\
  Polyvinylacetate & 130, 95, 78 & \cite{huang02,alba04,roland05} & 130, 61, 52 & \cite{huang02,alba04,roland05} & 0,43 & \cite{plazek91} \\
  Polystyrene & 77, 139 & \cite{huang02,plazek91} & 55 & \cite{huang02} & 0.35 & \cite{plazek91} \\
  Polymethylacrylate & 102, 122, 102 & \cite{huang02,roland04,plazek91}& 80, 94 & \cite{huang02,roland04} & 0.41 & \cite{plazek91} \\
  \hline
\end{tabular}
\caption{Fragilites and KWW stretching exponents of molecular liquids
  and polymers. The $^*$ indicates that the value is not given in the
  corresponding reference but is calculated from the data therein. The
  following abbreviations are used for the names of the liquids, PC = Propylene Carbonate,
BMPC = 1,1'-bis(p-methoxyphenyl)cyclohexane, BMMPC =
1,1'-di(4-methoxy-5-methylphenyl)cyclohexane, KDE = cresolphtalein-dimethyl-ether,
DEP A = diglycidylether of bisphenol A, and 
DHIQ = Decahydroisoquinoline.}
\end{table}
\end{landscape}

The correlation between stretching and fragility is not strikingly
different in figures \ref{fig:mP} and \ref{fig:mrho}. However, both on
theoretical ground (focusing on the intrinsic effect of temperature) and
on phenomenological one (isochoric fragility and stretching do not
appear to vary as one changes pressure along an isochrone), our
contention is that one should prefer using the isochoric fragility.

In the above we have considered only fragility and stretching at the
conventional glass transition temperature, that is around
$\tau_\alpha=100$ s.  However, we have pointed out in the Introduction
that both the steepness index characterizing fragility and the
stretching parameter depend on the relaxation time. Although still
debated, there seems to be a qualitative trend toward a decrease of
the stretching (an increase in $\beta_{KWW}$) and of the steepness
index as the relaxation time decreases and one approaches the
``normal'' liquid regime. It would certainly be very valuable to
obtain more data in order to study how the correlation between
fragility and stretching evolves as a function of the relaxation time.

\section{Conclusion}

In this article we have stressed the constraints that one should put
on the search for (meaningful) empirical correlations between the
fragility of a glassformer, which characterizes the temperature
dependence of the slowing down, and other dynamic or thermodynamic
properties. Among such constraints is the check that the proposed
correlations, often established at Tg and at atmospheric pressure, are
robust when one changes the reference relaxation time (in place of the
characteristic of Tg) as well as when one varies the pressure under
isochronic conditions. Important also is the fact that fragility
depends on the thermodynamic path considered (constant pressure versus
constant density) and that, contrary to the isobaric fragility, the
isochoric one appears as an intrinsic property of the glassformer,
characterizing the pure effect of temperature.

We have reported dielectric relaxation spectra under pressure for two
molecular liquids, m-toluidine and DBP. We have combined these data
with the available thermodynamic data and analyzed the respective
effect of density and temperature on the dynamics.  Our results are
consistent with a general picture in which the isochoric fragility is
constant on an isochrone. The shape of the relaxation function, as
e.g. expressed by the stretching parameter $\beta_{KWW}$, has also
been found constant along isochrones.

We have finally discussed the possible correlation between fragility
and stretching, suggesting that a meaningful correlation is to be
looked for between stretching and isochoric fragility, as both seem to
be constant under isochronic conditions and thereby reflect the
intrinsic effect of temperature. On the practical side, the
correlation is however no stronger with the isochoric fragility than
with the isobaric one. One top of large error bars that may be present
and that we have addressed in some detail, this reflects the fact that
correlations are rather statistical in nature, emerging from a
comparison of a large number of glassformers, rather than one-to-one
correspondences between properties of the materials. .

\section{Acknowledgment}
We would like to thank A. W\"urflinger for the PVT data on m-Toluidine
and Albena Nielsen and coworkers for making available her dielectric data on DBP
prior to publishing. We are grateful to Denis L'H\^ote and Fran{\c
  c}ois Ladieu for having lent us the SR830 lockin. Moreover we acknowledge
the work of Jo\"el Jaffr\'e who built the autoclave for the dielectric
measurements under pressure. This work was supported by the CNRS
(France) and grant No. 645-03-0230 from Forskeruddannelsesraadet
(Denmark).

\hspace{1cm}

\appendix
\section{Details on the density calculation}\label{sec:densAp}

The pressure and temperature dependences of the density are of course
a crucial input to the scaling shown in section \ref{sec:relaxDBP}. In
order to evaluate the effect of the extrapolations we have performed,
we focus on the scaling for the high-pressure room-temperature data of
Paluch and the data at atmospheric pressure, because the extrapolation of
the density is smallest in these cases. The discrepancies seen in
figure \ref{fig:dbpScaling} could be accounted for, if the density at
high pressure and room temperature were higher than what we have
estimated or if the density at low-temperature were lower than what we
have estimated.  The high-density dynamical data are taken at room
temperature. The experimental density data are also taken at room
temperature and they are only 
extrapolated above 1.2 GPa.  If the actual density is higher than
what we have estimated, then it means that the compressibility is larger
than what we taken. However, the compressibility at 1.2 GPa is
already in the high-pressure domain where it is very low and almost
pressure independent (it is slightly decreasing with increasing pressure).
The most conservative estimate we could make is to keep the
compressibility constant for pressures above the last experimental
point at 1.2 GPa. Such an approach changes
the ratio $\rho^{2.5}/T$ by less than one percent, and, therefore, can not account
for the discrepancy seen in figure \ref{fig:dbpScaling}. An alternative
explanation would be that the actual low temperature density is higher than
we have estimated, meaning that we have overestimated the expansion
coefficient $\alpha_P$. This latter has been calculated at
two different high temperatures based on the data in reference
\cite{bridgman32}. This leads to a slight decrease in expansion
coefficient with decreasing temperature. If the expansion coefficient
is to be smaller than the estimate from this temperature dependence,
then it would mean that the temperature dependence of the expansion
coefficient should increase as temperature decreases. This is the
opposite of what is seen in real liquids, where $\alpha_P$ at
atmospheric pressure tends to a constant at low temperatures
\cite{terminassian88}. It is actually most common to assume that the $\alpha_P$
of molecular liquids is constant below room temperature (e.g. ref.
\cite{reiser05}). This type of assumption would enhance the
discrepancy in figure \ref{fig:dbpScaling}. We therefore conclude that
the absence of collapse of the high-pressure data in
Fig. \ref{fig:dbpScaling} using a simple power law form for $e(\rho)$
cannot be explained by errors made in the estimating the PVT data.  

\section{Characterizing the spectral shape}\label{sec:shape}

In the following, we shortly review the procedures commonly used to characterize the shape of the relaxation spectrum of
viscous liquids and
test different descriptions on one of our spectra. We more
specifically look at schemes for converting one type of description to another.
This analysis is important for the present work because we compile
literature data in section \ref{sec:betam} in order to look at possible general
connections between relaxation shape and temperature dependence of the
relaxation time. 

The (normalized) Kohlrausch-William-Watts function or stretched exponential,
$\phi_{KWW} (t)=exp\left[-\(\frac{t}{\tau}\)^{\beta_{KWW}}\right]$,
leads to a loss peak in the frequency domain that is given by the
one-side Fourier transform

\begin{eqnarray}
  \phi{\prime\prime}_{KWW}(\omega)=\int_0^\infty-\diff{\phi_{KWW}(t) }{t}
 \;\sin(\omega t) \mathrm{d} t
\end{eqnarray}

The low-frequency behavior of this function is always a power law with
exponent 1. The high frequency behavior is a power law with exponent
$-\beta_{KWW}$ \cite{lindsey80}. $\beta_{KWW}$ is the only parameter
describing the shape of the relaxation function. Hence it controls
both  the exponent of the high frequency power law and the width of the relaxation function. 

The Havriliak-Negami (HN) function,
\begin{eqnarray}
  \label{eq:hn}
  \phi_{HN}(\omega)=\frac{1}{[1+(i\omega\tau_{HN})^\alpha]^\gamma}\;\;,
\end{eqnarray}
gives a power law with exponent $(-\alpha\gamma)$ in the high-frequency
limit and a power law of exponent $\alpha$ in the low frequency-limit of
its imaginary part.

The HN function reduces to Cole-Davidson (CD) one when $\alpha=1$. (In the case
of a CD function we follow the convention and refer to the $\gamma$
above as $\beta_{CD}$.) The CD spectrum has the same general
characteristics as the KWW one:
 a high-frequency power law with exponent given by
$\beta_{CD}$ and a low-frequency power law with exponent one.  However, the shape of
the two functions is not the same. The CD function is narrower
for a given high frequency exponent (given $\beta$) than the KWW function.
The best overall correspondence between the CD-function and the
KWW function has been determined by Lindsey and Patterson
\cite{lindsey80}.

No good correspondence exists in general between the HN and the KWW
functions. First of all because the former involves two adjustable shape
parameters and the latter only one (plus in both cases a parameter for
the intensity and one for the time scale). The KWW function always has a
slope of one at low frequencies while the HN function has a generally
nontrivial $\alpha$.  Alvarez \emph{et al.} \cite{alvarez91} numerically
found that the two functions can nonetheless be put in correspondence
by fixing the relation between the two HN parameters
$\gamma=1-0.812(1-\alpha)^{0.387}$ and choosing $\beta_{KWW}=(\alpha\gamma)^{(1/1.23)}$.
This restricted version of the HN function is sometimes referred to as
the AAC function \cite{gomez01}. The shape is described by one
parameter.  However, it is clear that this function cannot correspond
to the KWW function in the frequency range where the loss can be described by
power laws, as it was also noted by Gomez and
Alegria \cite{gomez01}. The AAC function inherits the behavior of the
HN function; as a result it has a nontrivial exponent $\alpha$ at low
frequencies and an exponent $-\alpha\gamma$ at high frequencies, while
the associated KWW function has exponents one and
$-\beta_{KWW}=-(\alpha\gamma)^{(1/1.23)}$ at low and high frequencies,
respectively.

Another approach is to describe the dielectric spectrum by a distribution of Debye
relaxations 
\begin{equation}
  \label{eq:dist}
  (\epsilon (\omega)-\epsilon_\infty)/\Delta \epsilon =\int_{-\infty}^\infty
  D(ln \tau)\frac{1}{1+i\omega \tau} \mathrm{d} ln \tau,
\end{equation}
and to fit the shape of the distribution $D(ln \tau)$ rather than the
spectral shape directly. The following form has been suggested for
the distribution function \cite{blochowicz03},
\begin{equation}
  \label{eq:gamma}
  D(ln \tau)=N \exp\( -\beta/\alpha
  (\tau/\tau_0)^\alpha\)(\tau/\tau_0)^\beta \(1+ \(\frac{\tau\sigma}{\tau_0}\)^{\gamma-\beta} \),
\end{equation}
where $N$ is a normalization factor. The function above is known as
the extended generalized gamma distribution, GGE. The last term (and
the parameters $\gamma$ and $\sigma$) describes a high-frequency wing,
corresponding to a change from one power law behavior (-$\beta$) to another
(-$\gamma$). This term can therefore be omitted if no wing is observed
in the spectrum. This results in a simpler distribution; the generalized gamma
distribution (GG) whose shape is described by two
parameters: $\alpha$ determines the width and $\beta$ gives the
exponent of the high-frequency power law. The low frequency is always
a power law with exponent one.

Finally, it is possible to describe the spectra phenomenologically in
terms of the full width at half maximum, usually normalized to the
full width at half maximum of a Debye peak \cite{dixon90} ($W/W_D$, with 
$W_D=1.14$ decade), and by the exponent of the power law describing
the high-frequency side. The power law exponent is not always well
defined, as there can be a high-frequency wing or a secondary process
appearing at high-frequencies.  Olsen \cite{olsen01} \emph{et al.}
therefore suggest to characterize the alpha peak by the minimal slope
found in a double logarithmic plot of the dielectric loss as a
function of frequency.  Note that this phenomenological description
requires two parameters to describe the shape, while the commonly used CD and the KWW
functions use only one parameter to describe the spectrum.

\begin{figure}[htbp]
  \centering
a) \includegraphics[scale=0.4]{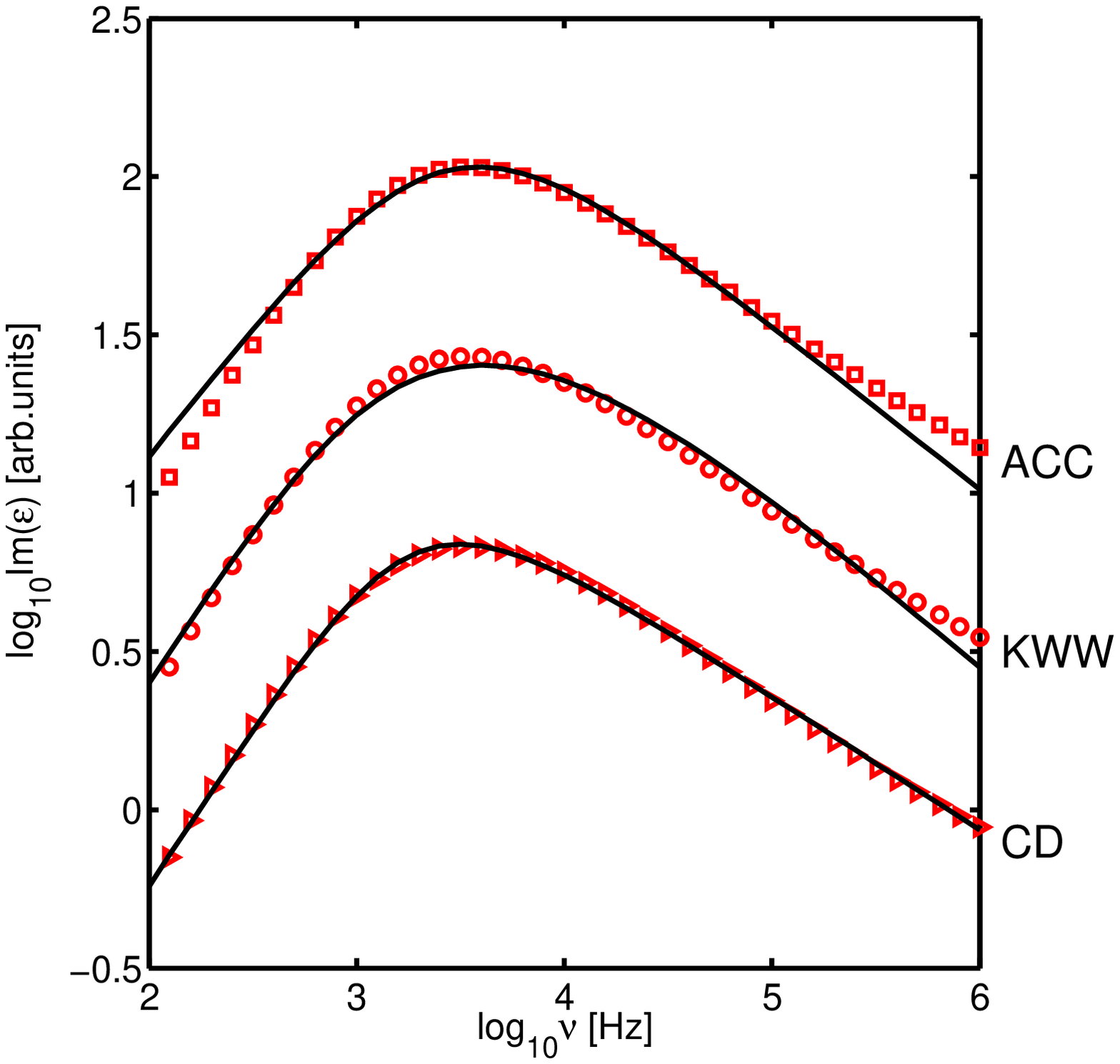}
b) \includegraphics[scale=0.4]{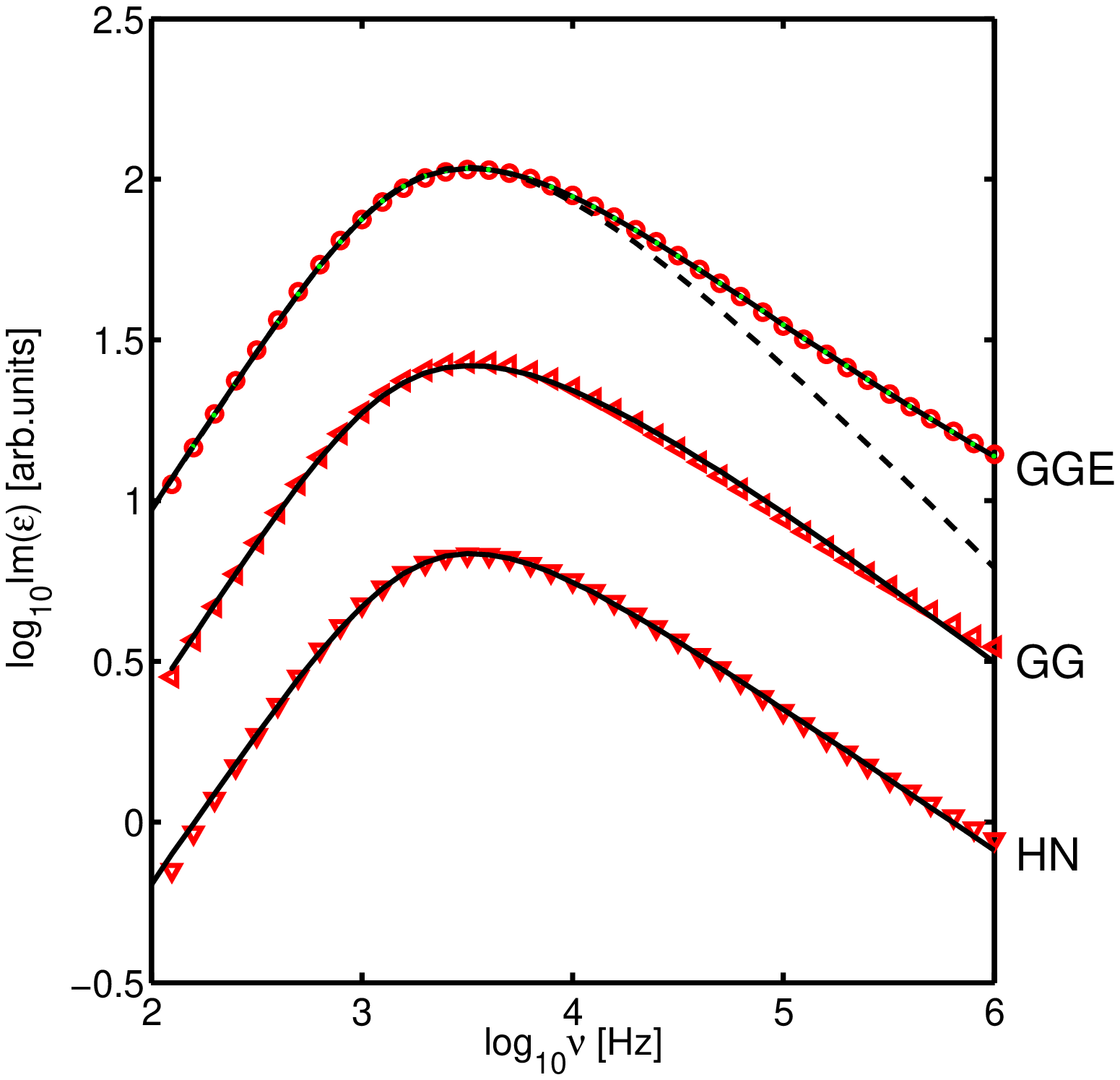}
  \caption{Log-log plot of the dielectric loss of m-toluidine at T=216.4K and
    122MPa along with best fits to several common functional forms.
 Figure a) show the fitting functions 
  from below and up; CD, KWW, AAC. Figure a) shows from below and up; HN, Gamma distribution,
    Generalized gamma distribution. CD, KWW and AAC have 1 parameter
    characterizing the shape, HN and Gamma have 2, and Generalized
    gamma has been fitted using 3 adjustable parameters. The dashed line
    shows the Gamma distribution corresponding to the generalized
    gamma distribution. The curves are displaced along the y axis by
    regular amounts.}\label{fig:mtolfit}
  \end{figure}

In figure \ref{fig:mtolfit} we show one of the dielectric spectra of
m-Toluidine along with fits to the functions described above. The
minimal slope is $-0.44$ and $W/W_D=1.56$.  The best fits to the
different functions are displayed in figure \ref{fig:mtolfit}. The CD-fit
gives $\beta_{CD}=0.42$, which with the Lindsey-Patterson scheme 
\cite{lindsey80} corresponds to $\beta_{KWW}\approx 0.55$. The direct
fit with the Fourier transform of the KWW gives $\beta_{KWW}=0.57$.
The best AAC fit gives $\alpha=0.85$ leading to $\gamma=0.61$ and
$\beta_{KWW}\approx(\gamma \alpha )^{1/1.23}=0.59$. This shows that
both the Patterson and the AAC approximations reasonably well
reproduce the $\beta_{KWW}$ value found from using KWW directly.
Another point worth noticing is that the $\beta_{KWW}$ value does not
correspond to the actual high-frequency slope. This is because the
overall agreement between the fit and the data is much more governed
by the width of the relaxation function than by its high-frequency
slope, as it is also clearly seen for the KWW fit in figure
\ref{fig:mtolfit}.  Note that the AAC approximation for the relation
between the HN parameters and $\beta_{KWW}$ only holds when the HN
parameters are fixed according to $\gamma=1-0.812(1-\alpha)^{0.387}$.
The original HN function has two adjustable parameters to describe the
shape. The best HN fit gives $\alpha=0.95$, and $\gamma=0.46$. The
Gamma distribution which also has two free parameters gives
$\alpha=40$ and $\beta=0.49$. Finally we have fitted with the GGE
using the constraint $\beta=3\gamma$ (see reference
\cite{blochowicz06}), meaning that the function has 3 free
parameters to discribe the shape, the values being $\alpha=40$,
$\beta=0.7$ $\sigma=53$ and $\gamma=\beta/3=0.23$. It is not
surprising that the GGE with 3 free parameters gives by far the best
fit. However it is also striking that the CD with only one parameter
describing the shape gives a good fit over the whole peak, whereas this
is not true for the KWW nor for the AAC.

From the above we conclude the CD-function gives a good description of
the shape of the relaxation using only one parameter to describe the
shape. We therefore use this function to fit our data. The KWW
exponent, $\beta_{KWW}$, does not give a proper measure of the high
frequency slope, but that it does give a reasonable one-parameter
measure of the overall shape of the dispersion. The KWW function is moreover the
function most commonly used in literature, which is the main reason
for using it in the discussion (section \ref{sec:disc}). 

\end{document}